\documentclass[aps,pra,twocolumn,superscriptaddress,10pt]{revtex4-2}

\usepackage{graphicx}
\usepackage{epstopdf}

\usepackage{dcolumn}
\usepackage{bm}
\usepackage{amsmath}
\usepackage{amssymb}
\usepackage{mathtools}
\usepackage{dsfont}
\usepackage{maybemath}
\usepackage{xcolor}
\usepackage{enumitem}
\usepackage{graphicx}

\usepackage{physics}
\usepackage{mathrsfs}
\usepackage{soul}
\DeclareUnicodeCharacter{2212}{-}

\newcommand{\sandw}[3]{\langle #1 | #2 | #3 \rangle}%
\newcommand{\stat}[1]{\langle #1 \rangle}%
\newcommand{\rarr}[0]{\rightarrow}

\newcommand{\eps}[0]{\varepsilon}

\newcommand{\up}[0]{\uparrow}
\newcommand{\dn}[0]{\downarrow}
\newcommand{\dw}[0]{\downarrow}
\newcommand{\ho}[0]{\mathrm{ho}}
\newcommand{\lu}[0]{\mathrm{lu}}
\newcommand{\lp}[0]{\left}
\newcommand{\rp}[0]{\right}
\newcommand{\s}[0]{\sigma}
\newcommand{\sbar}[0]{{\bar{\sigma}}}
\newcommand{\de}[0]{\partial}
\newcommand{\KS}[0]{\mathrm{KS}}

\newcommand{\tot}{\textrm{tot}}
\newcommand{\leqs}[0]{\leqslant}
\newcommand{\geqs}[0]{\geqslant}

\newcommand{\half}[0]{\frac{1}{2}}
\newcommand{\thalf}[0]{\tfrac{1}{2}}
\newcommand{\rr}[0]{\mathbf{r}}

\newcommand{\rrp}[0]{\mathbf{r}'}

\newcommand{\m}[0]{\textrm{min}}
\newcommand{\ens}{\textrm{ens}}

\newcommand{\defeq}{\vcentcolon=}
\newcommand{\eqdef}{=\vcentcolon}

\newcommand{\blue}[1]{\textcolor{black}{#1}}

\newcommand\identity{1\kern-0.25em\text{l}}

\newcommand{\kB}{k_{B}}

\newcommand{\NM}{{N\!M}}

\newcommand{\inn}{\mathrm{in}}
\newcommand{\out}{\mathrm{out}}
\begin{document}
\title{Many-electron systems with fractional electron number and spin: \\ exact properties above and below the equilibrium total spin value}

\author{Yuli Goshen}
\affiliation{Fritz Haber Center for Molecular Dynamics and Institute of Chemistry, The Hebrew University of Jerusalem, 9091401 Jerusalem, Israel}

\author{Eli Kraisler}
\email{Author to whom correspondence should be addressed: eli.kraisler@mail.huji.ac.il}
\affiliation{Fritz Haber Center for Molecular Dynamics and Institute of Chemistry, The Hebrew University of Jerusalem, 9091401 Jerusalem, Israel}

\date{\today}

\begin{abstract}

Description of many-electron systems with a fractional electron number, $N_\tot$, and fractional $z$-projection of the spin, $M_\tot$, is of great importance in physical chemistry, solid-state physics and materials science. In this work, we analyze the fundamental question of what is the ensemble ground state of a general, finite, many-electron system at zero temperature, with a given $N_\tot$ and $M_\tot$, distinguishing between low- and high-spin cases (separated by the boundary spin $M_B$). For the low-spin case, the general form of the ensemble ground state has been rigorously derived in J.\ Phys.\ Chem.\ Lett.\ \textbf{15}, 2337 (2024), generalizing the piecewise-linearity and the flat-plane conditions for many-electron systems.  Here we provide an alternative proof for this case, discuss the ambiguity in the description of the ground state and show that this ambiguity can be removed via maximization of the system's entropy.  For the high-spin case, we find that the form of the ensemble ground state strongly depends on the system in question. We prove three general properties, which characterize the ground state at high spins and narrow down the list of pure states it may consist of. We illustrate the aforementioned properties of high-spin cases by examining the ensemble ground state  when $M_\tot$ approaches $M_B$ from above, during addition of (a fraction of) an up- and down-electron to a given system. Furthermore, we relate the frontier orbital energies of Kohn-Sham (KS) density functional theory (DFT) to total energy differences at high spin values, particularly the ionization potential (IP), the fundamental gap and the spin flip energies. Analyzing the frontier energies on both sides of each boundary in the total energy profile, where the energy slope changes abruptly, we derive expressions for new derivative discontinuities, which are predicted to appear as jumps in the corresponding KS potentials. In this way, we generalize the well-known IP theorem of DFT to cases with fractional electron number and to cases with high spin. Our analytical results are supported by an extensive numerical analysis of the Atomic Spectra Database of the National Institute of Standards.
The new exact conditions for many-electron systems derived in this work are instrumental for development of advanced approximations in DFT and other many-electron methods.

\end{abstract}

\maketitle

\section{Introduction}
Simulation of physical and chemical processes in materials strongly relies on the ability to accurately obtain fundamental material properties from first-principles calculations~\cite{Szabo, Martin, Cramer04, Kohanoff06, ShollSteckel11, Giustino14_materials, Reining_MB, Marzari21}. One theoretical framework that is extremely popular in materials research is density functional theory (DFT)~\cite{DG,PY,Primer,EngelDreizler11}. The accuracy of a DFT calculation critically depends on the quality of the main approximation within the theory: that of the exchange-correlation (xc) functional. One way to create predictive xc approximations from first principles is by satisfying exact properties of many-electron systems~\cite{Perdew05,Perdew09,KaplanLevyPerdew23}. The idea relies on the fact that while the exact xc functional is not known, certain formal properties of many-electron systems are. Developing xc approximations that incorporate known properties by construction is expected to yield high accuracy together with a wide range of applicability. On a more general note, while the approach of satisfying exact constraints is most popular in DFT, it is also relevant within other quantum approaches to the many-electron problem~\cite{Szabo, FetterWalecka, Mattuck, Reining_MB}.

One important exact property of many-electron systems is the piecewise-linear behavior in ensemble states: For systems with a varying, possibly fractional total number of electrons, $N_\tot$, the ground state at zero temperature is described by the two-state ensemble $\hat{\Lambda} = (1-\alpha) \ketbra{\Psi_{N_0}} + \alpha \ketbra{\Psi_{N_0+1}}$, namely by a linear combination of the ground states with $N_0$ and $N_0+1$ electrons. Here $N_0 = \textrm{floor}(N_\tot)$ and $\alpha = N_\tot - N_0$ are, respectively, the integer and the fractional parts of the electron number $N_\tot$. As a result, the total energy of the system, the ensemble electron density and any property that is described by an operator, are piecewise-linear in~$\alpha$, i.e., in $N_\tot$~\cite{PPLB82,Yang00}. Particularly, the total energy of the system is linear between any two adjacent integer values of $N_\tot$, with the slope being the negative of the appropriate ionization potential (IP). As $N_\tot$ crosses an integer, the slope value is discontinuous; the change in the slope equals the fundamental gap.
Enforcing piecewise-linearity in an approximate functional -- e.g., by tuning a parameter~\cite{Stein10,Kronik_JCTC_review12,Sai11,Atalla16}, by an ensemble generalization~\cite{KraislerKronik13} or by developing a new functional~\cite{Zheng11,Dabo10,Dabo14}
-- can have an enormous, positive effect for the prediction of the IP via the highest occupied KS eigenvalue, for the fundamental gap of finite systems, for the description of dissociation, charge transfer, excitations and other properties (see, e.g.,~\cite{Teale08,Stein10,Refaely11,Kronik_JCTC_review12,GouldDobson13,KraislerKronik14,Borghi14,LiZhengYang15,KraislerKronik15,KraislerSchmidt15,Li18,NguyenColonna18,GouldKronikPittalis18,KraislerSchild20,KronikKummel20,Gould22,LavieGoshenKraisler23,Kraisler25} and references therein).

Accounting for the spin of electrons and investigating the energy dependence both on the total number of electrons, $N_\tot$, and on the $z$-projection of the spin, $M_\tot$~\cite{Chan99,CohenMoriSYang08, MoriS09, GalAyersProftGeerlings09, GalGeerlings10JCP, GalGeerlings10PRA, Cohen12, Saavedra12, GouldDobson13, XDYang16}, the constancy condition for the energy has been formulated~\cite{CohenMoriSYang08}. It states that the exact total energy must be constant when varying  $M_\tot$ at a constant, integer $N_\tot$, in the interval $[-S_\m, S_\m]$. Here $S_\m$ is the \emph{equilibrium} value of the \emph{total} spin, $S$, i.e., that value of the total spin $S$ for which the system has the lowest energy. Next, building upon previous results, the flat-plane condition for the energy has been derived~\cite{MoriS09}. Focusing on the H atom, Ref.~\cite{MoriS09} described its exact total energy, while continuously varying $N_\tot$ from 0 to 2 and $M_\tot$ accordingly, while $M_\tot \in [-\tfrac{1}{2}, \tfrac{1}{2}]$
~\footnote{Hartree atomic units are used throughout}.
The exact energy profile of H in the $N_\up - N_\dw$ plane has a diamond-like structure: it is comprised of two triangles whose vertices lie at the points of integer $N_\up$ and $N_\dw$.

In our recent Letter~\cite{GoshenKraisler24}, we
generalized the piecewise-linearity principle~\cite{PPLB82} and the flat-plane condition~\cite{MoriS09} to provide an exact description of the zero-temperature ensemble ground state of a general, finite, many-electron system.  We focused on the important, yet particular case of
\begin{equation}\label{eq:spin condition}
    |M_\tot| \leqs M_B,
\end{equation}
where $M_B$ is termed the boundary spin value, and equals
\begin{equation} \label{eq:MB_def}
    M_B = (1-\alpha) S_0 + \alpha S_1.
\end{equation}
Here $S_0$ and $S_1$ are the equilibrium spin values for systems with $N_0$ and $N_0+1$ electrons, respectively.

\begin{figure}
    \centering
    \includegraphics[width=0.98\linewidth]{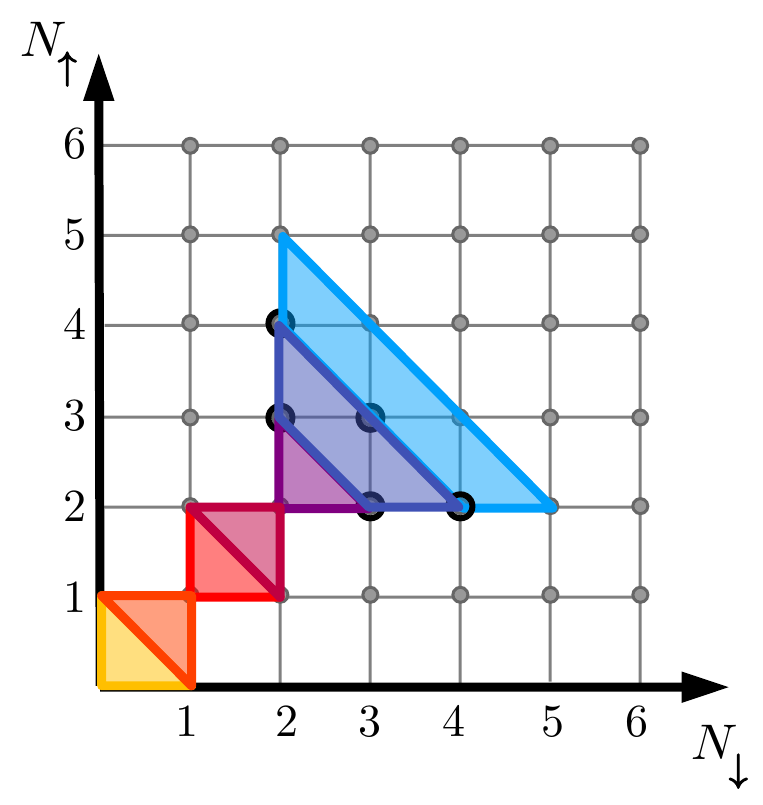}
    \caption{Illustration of Region~\eqref{eq:spin condition}, for the C atom with varying $N_\tot$ and $M_\tot$. The full gray circles correspond to pure states with integer $N_\up$ and $N_\dw$. The cyan region corresponds to $N_\tot \in [6,7]$ (i.e., C $\rarr$ C$^-$), the blue region -- to $N_\tot \in [5,6]$ (i.e., C$^+$ $\rarr$ C), and so on.
    The pure states that contribute to the ensemble ground state for $N_\tot \in [5,6]$ are emphasized with black circles.}
    \label{fig:trapezoids}
\end{figure}

Within Region~\eqref{eq:spin condition}, depicted in Fig.~\ref{fig:trapezoids}, and referred to here and below as the region \emph{within the trapezoids}, it has been proved in Ref.~\cite{GoshenKraisler24} which pure states contribute to the ground-state ensemble, and which do not. Then, using the exact ensemble ground state, the dependence of the energy and the spin-densities on both $N_\tot$ and $M_\tot$ has been characterized. A new derivative discontinuity, which manifests for spin variation at constant $N_\tot$ as a jump in the Kohn-Sham (KS) potential in DFT has been found. This phenomenon has been recently observed numerically in Ref.~\cite{Hayman25}.
Finally, a previously unknown degeneracy of the ground state has been identified, showing that the spin-densities $n_\s(\rr)$ may not be uniquely determined by setting the values of $M_\tot$ and $N_\tot$ alone; on the other hand, the total electron density $n(\rr)=n_\up(\rr)+n_\dw(\rr)$ is always known unambiguously.

In parallel, Ref.~\cite{BurgessLinscottORegan24} described the ground-state energy profile $E_\tot(N_{\tot},M_{\tot})$, also outside the trapezoids of Region~\eqref{eq:spin condition}, relying on the infinite-separation technique~\cite{Yang00}. It has been shown that the energy profile satisfies the tilted-plane condition, which gives rise to derivative discontinuities (see also~\cite{CapVigUll10,GoshenKraisler24}). Outside the trapezoids, a diverse set of tilted-plane structures for the energy has been discovered, relying on available  reference data~\cite{NIST_ASD}.

In the present article, we analyze the ensemble ground state of a general, finite, many-electron system at zero temperature, with a given $N_\tot$ and $M_\tot$, both for low- and high-spin cases, namely \emph{within} and \emph{beyond} Region~\eqref{eq:spin condition}.  In Sec.~\ref{sec:in_Reg_1}, for cases within Region~\eqref{eq:spin condition}, we provide an alternative proof for the ground state ensemble form, discuss the ambiguity in the description of the ground state, exemplify it by analyzing the ensemble spin polarization $\zeta(\rr)$ and suggest a new way to remove the aforementioned ambiguity: via maximization of the system's entropy.  For cases outside Region~\eqref{eq:spin condition}, in Sec.~\ref{sec:Beyond_Reg_1}, we find that the form of the ensemble ground state strongly depends on the system in question. We succeeded to prove three general properties for these cases, which characterize the ground state at high spins and narrow down the list of pure states it may consist of. In Sec.~\ref{sec:CasesPsiStar}, we illustrate the aforementioned properties of high-spin cases by examining the ensemble ground state when $M_\tot$ approaches $M_B$ from above, during addition of (a fraction of) an up- and down-electron to a given system. 
Next, in Sec.~\ref{sec:eigenvalues}, we relate the frontier orbital energies of Kohn-Sham (KS) density functional theory (DFT) to total energy differences at high spin values, particularly the ionization potentials (IPs) and the spin flip energies. Analyzing the frontier energies on both sides of each boundary in the total energy profile, where the energy slope changes abruptly, we derive expressions for new derivative discontinuities, which are predicted to appear as jumps in the corresponding KS potentials. In this way, we generalize the well-known IP theorem of DFT to be applicable to systems with high spin. Our analytical results are supported by an extensive numerical analysis of reference data available at the
Atomic Spectra Database (ASD) of the National Institute of Standards (NIST)~\cite{NIST_ASD}, performed in Sec.~\ref{sec:NumInvestigation}.  Section~\ref{sec:Conculsions} concludes the article.


\section{Theory}

\subsection{Ground state within Region~\eqref{eq:spin condition}} \label{sec:in_Reg_1}

In this Section, we provide an alternative derivation of the central proof of Ref.~\cite{GoshenKraisler24}, and determine the exact form of the ensemble ground state of a general, finite, many-electron system at zero temperature, within Region~\eqref{eq:spin condition}. We then discuss several outcomes of this proof, particularly focusing on the ambiguity in the ensemble ground state and ways to remove it.

\noindent \textbf{Definitions.} Consider a system with a varying, possibly fractional, number of $\up$- and $\dw$-electrons, denoted $N_\up$ and $N_\dw$, respectively. The total number of electrons, $N_\tot = N_\up + N_\dw$, is expressed as $N_\tot = N_0 + \alpha$,  where $N_0 = \textrm{floor}(N_\tot) \in \mathbb{N}_0$ and $\alpha \in [0,1)$, are the integer and fractional parts of $N_\tot$, respectively. The total $z$-projection of the spin is $M_\tot = \thalf (N_\up - N_\dw)$. Since $N_\tot$ and $M_\tot$ can be fractional, the ground state of such a system cannot always be described by a pure state. Instead, it is described by an ensemble, which generally can be expressed as
\begin{equation} \label{eq:Lambda_def}
    \hat{\Lambda} = \sum_{N=0}^{N_\mathrm{max}} \sum_{M = -N/2}^{N/2}  \lambda_{N,M} \ket{\Psi_{N,M}} \bra{\Psi_{N,M}}.
\end{equation}
Here $\ket{\Psi_{N,M}}$ is the pure ground state for a system with $N$ electrons, where the $z$-projection of the spin is $M$. The summation over $N$ goes from 0 (a state with no electrons) to $N_\mathrm{max}$ -- the maximal electron number a given system can bind. The inner summation over $M$ goes from $-N/2$ (a fully down-polarized state, where $N_\up = 0$ and $N_\dw=N$) to $N/2$ (a fully up-polarized state, where $N_\up = N$ and $N_\dw=0$). If $N$ is even, $M$ is an integer and the summation over $M$ goes over integers. If $N$ is odd, $M$ is a half-integer, and the summation over $M$ goes over half-integers, with an interval of 1.

$\{ \lambda_\NM \}$ are the statistical weights of the various states that comprise the ensemble. As a direct consequence,
\begin{equation} \label{eq:lambda_0_1}
    \lambda_\NM  \in [0,1]  \:\:\:  \forall N, M
\end{equation}
and
\begin{equation} \label{eq:Tr_Lambda}
    \sum_{N=0}^{N_\mathrm{max}} \sum_{M = -N/2}^{N/2}  \lambda_\NM = \Tr \{ \hat{\Lambda} \} = 1.
\end{equation}

Using Eq.~\eqref{eq:Lambda_def}, the expectation value of a quantity associated with an operator $\hat{O}$ (which for a pure ground state \blue{$\ket{\Psi}$ would just be $O = \sandw{\Psi}{\hat{O}}{\Psi}$}), is expressed as
$O = \Tr \{ \hat{\Lambda} \hat{O} \} = \sum_{N=0}^{N_\mathrm{max}} \sum_{M = -N/2}^{N/2}  \lambda_\NM O_\NM$, where $O_\NM = \sandw{\Psi_\NM}{\hat{O}}{\Psi_\NM}$.

In particular, the total energy of the above ensemble ground state equals
\begin{equation} \label{eq:E_tot}
    E_\tot = \Tr \{ \hat{\Lambda} \hat{H} \} = \sum_{N=0}^{N_\mathrm{max}} \sum_{M = -N/2}^{N/2}  \lambda_\NM E_\NM,
\end{equation}
where $\hat{H}$ is the Hamiltonian, being usually defined as $\hat{H} =\hat{T}+\hat{V}+\hat{W}$, i.e.\ as the sum of the kinetic energy operator $\hat{T}$, the external potential operator $\hat{V}$, and the electron-electron repulsion operator $\hat{W}$.

We note that one can consider ensemble ground states that are more general than what Eq.~\eqref{eq:Lambda_def} suggests, namely $\sum_{i} D_i \ketbra{\chi_i}$, where $\ket{\chi_i}$ are any states, and they are linear combinations of the form \blue{$\ket{\chi_i}=\sum_\NM c_\NM^i \ket{\Psi_\NM}$}. Such general ensembles will contain off-diagonal terms $\ketbra{\Psi_{N_1,M_1}}{\Psi_{N_2,M_2}}$. However, as long as our Hamiltonian and other operators of interest commute with the electron number operator, $\hat{N}$, and the operator for the $z$-projection of the spin, $\hat{S}_z$, this generalization does not affect any expectation values, and therefore we consider only ensembles as in Eq.~\eqref{eq:Lambda_def}.

Our task is  to determine the weights $\{ \lambda_\NM \}$ such that the total energy of Eq.~\eqref{eq:E_tot} is minimized, while the following constraints are kept:

\noindent The total number of electrons,
\begin{equation} \label{eq:N_tot}
    N_\tot = \Tr \{ \hat{\Lambda} \hat{N} \} = \sum_{N=0}^{N_\mathrm{max}} \sum_{M = -N/2}^{N/2}  \lambda_\NM N = N_0 + \alpha,
\end{equation}
The total $z$-projection of the spin,
\begin{equation} \label{eq:M_tot}
    M_\tot = \Tr \{ \hat{\Lambda} \hat{S}_z \} = \sum_{N=0}^{N_\mathrm{max}} \sum_{M = -N/2}^{N/2}  \lambda_\NM M,
\end{equation}
as well as the probabilistic constraints outlined in Eqs.~\eqref{eq:lambda_0_1} and~\eqref{eq:Tr_Lambda}. Ref.~\cite{GoshenKraisler24} found which states $\ket{\Psi_\NM}$ contribute to the ensemble ground state and which do not (because their weights vanish), providing a rigorous mathematical proof by contradiction. Here we offer an alternative proof for the same statement.

\noindent \textbf{Derivation.} Notice that as long as $\hat{H}$ does not include an external magnetic field, $E_{N,-M} = E_{N,M}$. As a result, it is useful to define
\begin{equation} \label{eq:ell_def}
    \ell_{N,M} = \begin{cases}
			\lambda_{N,M} + \lambda_{N,-M}             & :  M \neq 0 \\
            \lambda_{N,0} & :  M=0
		 \end{cases} \:\:\: ,
\end{equation}
because the expectation value for any quantity that satisfies $O_{N,-M} = O_{N,M}$ can be expressed as
\begin{equation}
    O = \sum_{N=0}^{N_\mathrm{max}} \sum_{M  \geqs 0}  \ell_{N,M} O_{N,M},
\end{equation}
where $M$ starts from 0 for even $N$ and from $\thalf$ for odd $N$, and runs up to $N/2$.

We now introduce some additional information on the ground-state energies $E_\NM$. For a given, fixed $N$, eigenstates of $\hat{H}$ are characterized not only by $M$, but also by the quantum number $S$, such that $\hat{S}^2 \ket{\Psi_{N,S,M}} = S(S+1) \ket{\Psi_{N,S,M}}$. For each value of $S$, there are $(2S+1)$ corresponding values of $M$, namely $M = -S \dots S$, which form a multiplet. All states in the multiplet have the same energy. It is also assumed, for matters of simplicity, that a pure state with given $N$ and $M$ is not further degenerate.

The specific value of $S$ which corresponds to the lowest-energy multiplet is denoted here $S_\m$.
For all $M=- S_\m\dots S_\m$, the pure states $\ket{\Psi_\NM}$ belong to that multiplet, and their energies are $E_\NM=E_N$, the minimum energy of the $N$-electron system.
States with $|M| > S_{\m}$ belong to a different multiplet (i.e., a different value of $S$), and have a higher energy. 
$S_{\m}$ is a function of $N$: for example, for the neutral C atom ($N=6$), $S_{\m}(6)=1$, for its first cation $S_{\m}(5) = \thalf$, for the second cation $S_{\m}(4) = 0$, and so on. For further convenience, let us denote $S_0 = S_{\m}(N_0)$ and  $S_1 = S_{\m}(N_0+1)$.

Equipped with the information provided above, we now turn back to Eq.~\eqref{eq:E_tot} and express the total energy as
\begin{align}
E_\tot &= \sum_{N=0}^{N_\mathrm{max}} \sum_{M  \geqs 0}  \ell_{N,M} E_{N,M} = \nonumber \\
       &= E_{N_0,S_0} + \sum_{N=0}^{N_\mathrm{max}} \sum_{M  \geqs 0}  \ell_{N,M} (E_{N,M} - E_{N_0,S_0}),
\end{align}
where we singled out the ground-state energy for $N_0$ electrons, $E_{N_0,S_0}$, using Eq.~\eqref{eq:Tr_Lambda}. Next, we split the sum over $N$, explicitly writing out the terms with $N=N_0$ and $N=N_0+1$:
\begin{align} \label{eq:E_tot_interm1}
E_\tot &= E_{N_0,S_0} + \sum_{M  \geqs 0}  \ell_{N_0,M}   (E_{N_0,M} - E_{N_0,S_0})    \nonumber \\
&+ \sum_{M  \geqs 0}  \ell_{N_0+1,M} (E_{N_0+1,M} - E_{N_0,S_0})  \nonumber \\
&+ \sum_{N=0}^{N_0-1} \sum_{M  \geqs 0}  \ell_{N,M} (E_{N,M} - E_{N_0,S_0}) \nonumber \\
&+ \sum_{N=N_0+2}^{N_\mathrm{max}} \sum_{M  \geqs 0}  \ell_{N,M} (E_{N,M} - E_{N_0,S_0}),
\end{align}
Focusing on the second term on the right-hand side (rhs) of Eq.~\eqref{eq:E_tot_interm1}, we realize that terms with $M \leqs S_0$ vanish, because they all belong to the same multiplet and have the same energy, $E_{N_0,S_0}$.
To make progress with the third term, we denote $A_1 = - (E_{N_0+1,S_1} - E_{N_0,S_0})$. [If the system is electrically neutral for $N_0$ electrons, $A_1$ is nothing else but the (first) electron affinity, but no loss of generality takes place here].
We now split the sum over $M$ in the third term in two: up to $S_1$ and beyond $S_1$. For $M \leqs S_1$, we realize that $(E_{N_0+1,M} - E_{N_0,S_0}) = -A_1$. Thus, we obtain
\begin{align} \label{eq:E_tot_interm2}
& E_\tot = E_{N_0,S_0} + \sum_{M  > S_0}  \ell_{N_0,M}   (E_{N_0,M} - E_{N_0,S_0})    \nonumber \\
&- \sum_{M  \geqs 0}^{S_1}  \ell_{N_0+1,M} A_1 + \sum_{M  > S_1}  \ell_{N_0+1,M} (E_{N_0+1,M} - E_{N_0,S_0})  \nonumber \\
&+ \sum_{N=0}^{N_0-1} \sum_{M  \geqs 0}  \ell_{N,M} (E_{N,M} - E_{N_0,S_0}) \nonumber \\
&+ \sum_{N=N_0+2}^{N_\mathrm{max}} \sum_{M  \geqs 0}  \ell_{N,M} (E_{N,M} - E_{N_0,S_0}),
\end{align}
Next, we turn to Eq.~\eqref{eq:N_tot}, and using Eqs.~\eqref{eq:ell_def} and~\eqref{eq:Tr_Lambda}, express it as
\begin{equation}
N_\tot = \sum_{N=0}^{N_\mathrm{max}} \sum_{M  \geqs 0}  \ell_{N,M} N = N_0 + \sum_{N=0}^{N_\mathrm{max}} \sum_{M  \geqs 0}  \ell_{N,M} (N-N_0).
\end{equation}
We then realize that $\alpha = \sum_{N=0}^{N_\mathrm{max}} \sum_{M  \geqs 0}  \ell_{N,M} (N-N_0)$. Furthermore, if we single out the $(N_0+1)$-term and notice that the $N_0$-term vanishes, we get
\begin{align} \label{eq:alpha_lNM}
\alpha &= \sum_{M \geqs 0}  \ell_{N_0+1,M} + \sum_{N=0}^{N_0-1} \sum_{M  \geqs 0}  \ell_{N,M} (N-N_0) + \nonumber \\
& + \sum_{N=N_0+2}^{N_\mathrm{max}} \sum_{M  \geqs 0}  \ell_{N,M} (N-N_0).
\end{align}
Focusing on the first term on the rhs of the above equation, splitting the summation over $M$ up to $S_1$ and beyond $S_1$, and multiplying both sides by $A_1$, we find that
\begin{align}
&-\sum_{M \geqs 0}^{S_1}  \ell_{N_0+1,M} A_1 = -\alpha A_1 + \sum_{M > S_1}  \ell_{N_0+1,M} A_1 + \nonumber \\
&+ \sum_{N=0}^{N_0-1} \sum_{M  \geqs 0}  \ell_{N,M} (N-N_0) A_1 \nonumber \\
&+ \sum_{N=N_0+2}^{N_\mathrm{max}} \sum_{M  \geqs 0}  \ell_{N,M} (N-N_0) A_1.
\end{align}
Using this result in Eq.~\eqref{eq:E_tot_interm2} and recalling the definition of $A_1$, we arrive at the following form of the total energy:
\begin{align} \label{eq:E_tot_sq_brackets}
& E_\tot = (1-\alpha) E_{N_0,S_0} + \alpha E_{N_0+1,S_1} + \nonumber\\
& +\sum_{M  > S_0}  \ell_{N_0,M}   \lp[ E_{N_0,M} - E_{N_0,S_0} \rp]    \nonumber \\
& + \sum_{M  > S_1}  \ell_{N_0+1,M} [E_{N_0+1,M} - E_{N_0+1,S_1}]  \nonumber \\
& + \sum_{N=0}^{N_0-1} \sum_{M  \geqs 0}  \ell_{N,M} [E_{N,M} - E_{N_0,S_0} + (N-N_0)A_1] \nonumber \\
& + \sum_{N=N_0+2}^{N_\mathrm{max}} \sum_{M  \geqs 0}  \ell_{N,M} [E_{N,M} - E_{N_0,S_0} + (N-N_0)A_1].
\end{align}

Our task now is to show that all the expressions that appear in square brackets in Eq.~\eqref{eq:E_tot_sq_brackets} are strictly positive. For the first and second square brackets, this follows immediately: all states with $M>S_0$ yield an energy higher than $E_{N_0,S_0}$; a similar statement is true for $M>S_1$.

For the third and fourth brackets, we employ the \emph{convexity conjecture} for the energy~\cite{PPLB82,Lieb,DG,Cohen12,Kraisler_PhD, Ayers_convexity_24}. To this end, we first analyze the term
\begin{equation} \label{eq:interm_term}
E_{N,S_{\m}(N)} - E_{N_0,S_0} + (N-N_0)A_1.
\end{equation}
For $0 \leqs N \leqs N_0-1$, Expression~\eqref{eq:interm_term} is smaller or equal to the third square bracket of Eq.~\eqref{eq:E_tot_sq_brackets}, and if the former is positive, so is the latter. Similarly, for $N_0+2 \leqs N \leqs N_\mathrm{max}$, Expression~\eqref{eq:interm_term} is smaller or equal to the fourth square bracket of Eq.~\eqref{eq:E_tot_sq_brackets}.

To proceed, we define the $k^\mathrm{th}$ ionization energy as $I_k \defeq E_{N_0-k} - E_{N_0-k+1} \equiv E_{N_0-k,S_{\m}(N_0-k)} - E_{N_0-k+1,S_{\m}(N_0-k+1)}$, for all $1 \leqs k \leqs N_0$. Then, for $0 \leqs N \leqs N_0 − 1$, we have $E_N - E_{N_0} = (E_N - E_{N-1}) + (E_{N-1} - E_{N-2}) + \cdots + (E_{N_0-1} - E_{N_0}) = \sum_{k=1}^{N_0-N} I_k$. The convexity conjecture states that the energy sequence $\{ E_N \}$ is convex and monotonically decreasing with $N$. This means that all the ionization energies/electron affinities are positive, and higher ionizations are always larger than the lower ones: $ \cdots > I_2 > I_1 > A_1 > \cdots $. This conjecture, although strongly supported by experimental data~\cite{HandChemPhys92}, remains without proof, to the best of our knowledge~\cite{PPLB82,Lieb,DG,Cohen12,Ayers_convexity_24}; see also~\cite{Burgess_convexity_23} in this context. As a consequence of the convexity conjecture, for $0 \leqs N \leqs N_0 − 1$, we find that $E_{N,S_{\m}(N)} - E_{N_0,S_0} + (N-N_0)A_1 = E_N - E_{N_0} - (N_0-N) A_1 = \sum_{k=1}^{N_0-N} (I_k - A_1)>0$, because \emph{each} term in the sum above is positive.  Therefore, the third square brackets in Eq.~\eqref{eq:E_tot_sq_brackets} are positive.

For the fourth square brackets in Eq.~\eqref{eq:E_tot_sq_brackets}, we define the $k^\mathrm{th}$ electron affinity $A_k = E_{N_0+k-1} - E_{N_0+k}$, for all $1 \leqs k \leqs N_\mathrm{max} - N_0$
~\footnote{We stress again that there is no loss of generality: in case the state with $N_0$ electrons does not correspond to an electrically neutral system, some of the $A_k$'s or $I_k$'s should be renamed to ionization/affinity, respectively.}.~
Then, for $N_0+1 \leqs N \leqs N_\mathrm{max}$, we have $E_N - E_{N_0} = (E_N - E_{N-1}) + (E_{N-1} - E_{N-2}) + \cdots + (E_{N_0+1} - E_{N_0}) = - \sum_{k=1}^{N-N_0} A_k$. Consequently, $E_{N,S_{\m}(N)} - E_{N_0,S_0} + (N-N_0)A_1 = \sum_{k=1}^{N-N_0} (A_1 - A_k)$, which, given the convexity conjecture, is also positive.

We have therefore shown that in Eq.~\eqref{eq:E_tot_sq_brackets} all the terms that appear in square brackets are strictly positive. Taking into account the fact that the weights $\ell_{N,M}$ cannot be negative, the ensemble total energy $E_\tot$ is minimized if all the weights that multiply the
aforementioned bracketed terms are set to zero. This choice is possible for $M_\tot$ values within~\eqref{eq:spin condition} (see  discussion around Eq.~\eqref{eq:M_upper_bound} below). Therefore, $\ell_{N,M} = 0$ if $N \neq N_0$ or $N_0+1$, irrespective of $M$; for $N=N_0$, $\ell_{N_0,M} = 0$ for $M>S_0$; for $N=N_0+1$, $\ell_{N_0+1,M} = 0$ for $M>S_1$.
In terms of the weights $\{ \lambda_{N,M} \}$, \emph{the only terms} in the ensemble $\hat{\Lambda}$ of Eq.~\eqref{eq:Lambda_def} that survive the minimization are  $\ket{\Psi_{N_0,-S_0}} \bra{\Psi_{N_0,-S_0}}, \cdots, \ket{\Psi_{N_0,S_0}} \bra{\Psi_{N_0,S_0}}$ and $\ket{\Psi_{N_0+1,-S_1}} \bra{\Psi_{N_0+1,-S_1}}, \cdots, \ket{\Psi_{N_0+1,S_1}} \bra{\Psi_{N_0+1,S_1}}$. This concludes our derivation.

From Eq.~\eqref{eq:E_tot_sq_brackets} we now realize that
\begin{align} \label{eq:E_tot_PL}
E_\tot = (1-\alpha) E_{N_0} + \alpha E_{N_0+1},
\end{align}
i.e.\ that the total energy is piecewise-linear in $\alpha$.
Looking back at Eq.~\eqref{eq:alpha_lNM}, we find that $\alpha = \sum_{M \geqs 0}^{S_1} \ell_{N_0+1,M}$, and employing Eq.~\eqref{eq:Lambda_def} we realize that $\sum_{M \geqs 0}^{S_0} \ell_{N_0,M} = 1-\alpha$.

To summarize, the ground state of a system with $N=N_0+\alpha$ electrons with the $z$-projection of the spin being equal $M_\tot$, is described by the ground state
\begin{align} \label{eq:Lambda_gs}
    \hat{\Lambda} &= \sum_{M=-S_0}^{S_0} \lambda_{N_0,M} \ket{\Psi_{N_0,M}} \bra{\Psi_{N_0,M}} + \nonumber \\
    &+ \sum_{M=-S_1}^{S_1} \lambda_{N_0+1,M} \ket{\Psi_{N_0+1,M}} \bra{\Psi_{N_0+1,M}},
\end{align}
where the statistical weights $\{ \lambda_{N,M} \}$ obey the following three restrictions:
\begin{align}
    & \sum_{M = -S_0}^{S_0} \lambda_{N_0,M} = 1-\alpha \label{eq:lambda_1_m_alpha}\\
    & \sum_{M = -S_1}^{S_1} \lambda_{N_0+1,M} = \alpha \label{eq:lambda_alpha}\\
    & \sum_{M = -S_0}^{S_0} \lambda_{N_0,M} M + \sum_{M = -S_1}^{S_1} \lambda_{N_0+1,M} M  = M_\tot. \label{eq:lambda_Mtot}
\end{align}

Interestingly, throughout the derivation, the requirement on $M_\tot$ was not used. Furthermore, it may look as if $M_\tot$ can obtain any value. However, closely looking at Eqs.~\eqref{eq:lambda_1_m_alpha}--\eqref{eq:lambda_Mtot}, we see that $M_\tot$ is confined: the first sum in Eq.~\eqref{eq:lambda_Mtot} is bound from above by replacing $M$ with its maximal value:
\begin{equation} \label{eq:M_upper_bound}
    \sum_{M = -S_0}^{S_0} \lambda_{N_0,M} M \leqs \sum_{M = -S_0}^{S_0} \lambda_{N_0,M} S_0 = (1-\alpha) S_0
\end{equation}
This sum is also bound from below by replacing $M$ with its minimal value: $\sum_{M = -S_0}^{S_0} \lambda_{N_0,M} M \geqs -(1-\alpha) S_0$. The same logic applies also to the second sum in Eq.~\eqref{eq:lambda_Mtot}. Combining all these bounds, we realize that in our derivation $-M_B \leqs M_\tot \leqs M_B$, where
$M_B$ is given by Eq.~\eqref{eq:MB_def}.
In other words, taking all the $\ell_{N,M}$'s explicitly written in Eq.~\eqref{eq:E_tot_sq_brackets} to zero, without explicitly treating $M_\tot$, gives us a global minimum, for which $M_\tot$ is constrained as in Eq.~\eqref{eq:spin condition}. Treating the ensemble ground state of systems with $|M_\tot|>M_B$ requires  \blue{a different approach, as given in Sections~\ref{sec:Beyond_Reg_1} and onwards}.

\noindent \textbf{Outcomes: Piecewise-linearity.} The fact that the ground state can be expressed as in Eq.~\eqref{eq:Lambda_gs}, with Eqs.~\eqref{eq:lambda_1_m_alpha}, \eqref{eq:lambda_alpha} and~\eqref{eq:lambda_Mtot} as restrictions, has several direct outcomes.  One of them is that the total energy is piecewise-linear in $N_\tot$ and independent of $M_\tot$ (Eq.~\eqref{eq:E_tot_PL}). Importantly, the same behavior applies not only to the energy, but actually to a large family of quantities: Any expectation value of an operator $\hat{P}$,
which is commutative with the ladder operator for the spin, $\hat{S}_+ = \hat{S}_x + i \hat{S}_y$, is piecewise-linear in $\alpha$ and is independent of $M_\tot$, within Region~\eqref{eq:spin condition}:
\begin{equation}\label{eq:P__PL}
    P(N_\tot,M_\tot) = (1-\alpha) P_{N_0} + \alpha P_{N_0+1}.
\end{equation}

To prove this statement, let us show that for pure states, the expectation value $P_{N,M} = \expval{\hat{P}}{\Psi_{N,M}}$ is \emph{independent} of $M$, as long as $M \in [-S_\m(N), S_\m(N)]$.  First, recall that by definition, applying the spin ladder operators to a pure state yields
\begin{equation} \label{eq:S_pm}
\hat{S}_\pm \ket{\Psi_{N,M \mp 1}} = C^\pm_{N,M} \ket{\Psi_{N,M}}
\end{equation}
where $\hat{S}_- = \hat{S}_+^\dagger = \hat{S}_x - i \hat{S}_y$. It then follows that the state $ \ket{\Psi_{N,M}}$ is an eigenstate of the operator $\hat{S}_+ \hat{S}_-$.

Second, compare the expectation values $P_{N,M}$ and $P_{N,M-1} = \expval{\hat{P}}{\Psi_{N,M-1}}$.
Expressing $\ket{\Psi_{N,M-1}}$ and $\bra{\Psi_{N,M-1}}$ using Eq.~\eqref{eq:S_pm}, we readily understand that $P_{N,M-1}$ is proportional to $\sandw{\Psi_{N,M}}{\hat{S}_+\hat{P}\hat{S}_-}{\Psi_{N,M}}$. More precisely,
\begin{equation}\label{eq:P_N_Mminus1}
    P_{N,M-1} = \frac{\sandw{\Psi_{N,M}}{\hat{S}_+\hat{P}\hat{S}_-}{\Psi_{N,M}}}{\sandw{\Psi_{N,M}}{\hat{S}_+\hat{S}_-}{\Psi_{N,M}}}
\end{equation}
In case $[\hat{P}, \hat{S}_+] = 0$, we can interchange these two operators in the numerator. Then, acting with $\hat{S}_+ \hat{S}_-$ on $\ket{\Psi_{N,M}}$, which is an eigenstate of the latter operator, we release a constant; this is the same constant both in the numerator and the denominator. As a result, we find that $P_{N,M-1} = P_{N,M}$. In other words, $P_{N,M}$ is $M$-independent; therefore, it can be denoted simply as $P_N$ (which we already did in Eq.~\eqref{eq:P__PL}).

Finally, when calculating the ensemble expectation value of $\hat{P}$ via Eq.~\eqref{eq:Lambda_gs}, using the $M$-independence property we just established, we get
\begin{align} \label{eq:P__exp_val}
    P(N_\tot,M_\tot) = \Tr \{ \hat{\Lambda} \hat{P} \} =
    \nonumber \\
    = \sum_{M=-S_0}^{S_0} \lambda_{N_0,M} P_{N_0}
    &+ \sum_{M=-S_1}^{S_1} \lambda_{N_0+1,M} P_{N_0+1}.
\end{align}
By pulling $P_{N_0}$ and $P_{N_0+1}$ out of the sums and using Eqs.~\eqref{eq:lambda_1_m_alpha} and~\eqref{eq:lambda_alpha}, we arrive at the piecewise-linearity property of $P(N_\tot,M_\tot)$, i.e., Eq.~\eqref{eq:P__PL}.

Piecewise-linearity applies therefore to a variety of properties: Such is the total energy (in absence of a magnetic field, as discussed here), the electron density $n(\rr)$, the reduced density matrix, $\rho_1(\rr,\rrp)$, the Green function $G(\rr t, \rrp t)$ and any property that depends only on spatial, and not on spin coordinates. However, this is not the case for the energy in presence of a magnetic field, or for the spin-density $n_\s(\rr)$.

\noindent \textbf{Outcomes: Ground-state non-uniqueness.}
Another surprising outcome of Eq.~\eqref{eq:Lambda_gs} is the fact that the ensemble ground state may not be uniquely defined~\cite{GoshenKraisler24}. Indeed, the number of statistical weights $\lambda_\NM$ is $2(S_0+S_1+1)$, whereas the number of equalities that aim to determine them is three (Eqs.~\eqref{eq:lambda_1_m_alpha}, \eqref{eq:lambda_alpha} and~\eqref{eq:lambda_Mtot}); in addition, there also exist the inequalities of Eq.~\eqref{eq:lambda_0_1}.  This means that in order to fully determine the ground state of a many-electron system, it is not enough to specify its total electron number and spin, or, equivalently, the number of up and down electrons.

Cases where no ambiguity in the ground state occurs are: (i) When $S_0=0$ and $S_1 = \half$ and \textit{vice versa}, as the number $\lambda_\NM$'s is three; (ii) When $N_\up$ is made fractional, but $N_\dw$ is a constant integer and \textit{vice versa}, namely the common case of ionizing a system by removing an electron from a specific spin channel. There, the ensemble state consists of two states only, with uniquely determined coefficients:
\begin{equation}\label{eq:statement_0}
(1-\alpha) \ketbra{\Psi_{N_0,S_0}} + \alpha \ketbra{\Psi_{N_0+1,S_1}},
\end{equation}
with $S_1 = S_0 \pm \half$.
(iii) In any case, the total energy, the total density and any quantity $P$ as in Eq.~\eqref{eq:P__PL}, are always uniquely defined. Different is the situation for $n_\s(\rr)$. Therefore, we may have the situation where there is a multitude of spin-densities $n_\up(\rr)$ and $n_\dw(\rr)$, which sum up to the same total density and yield the same total energy. We illustrate this below, taking as an example the C atom with fractional electron number.

\begin{figure}
    \centering
\includegraphics[width=0.98\linewidth]{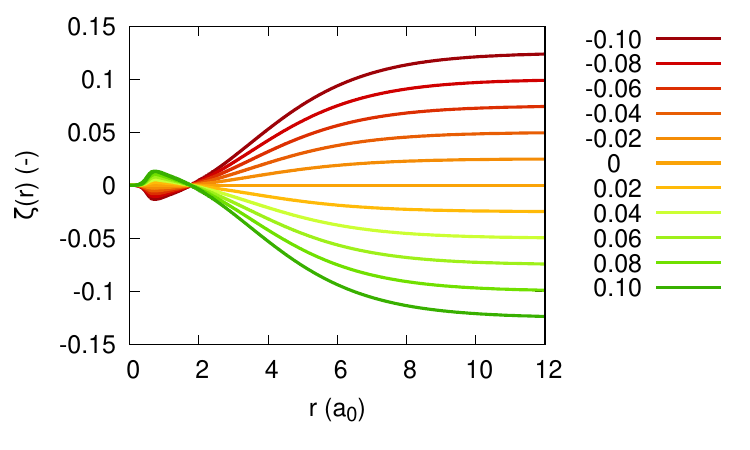}
\caption{Spin polarization $\zeta(r)$, for the C atom, with $\alpha=0.8$ and $M_\tot=0$, for various values of the parameter $y$ defined in the text (see Legend).}
    \label{fig:C__zeta}
\end{figure}

Consider, therefore, the case where $N_0 = 5$, $\alpha \in (0,1)$, $S_0 = \thalf$ and $S_1 = 1$. The ensemble in this case consists of five terms, in accordance with Eq.~\eqref{eq:Lambda_gs}, and its  coefficients can be expressed as $\lambda_{N_0, \pm \half} = \thalf(1-\alpha) \pm y$, $\lambda_{N_0+1, \pm 1} = \thalf(\alpha - x) \pm \thalf (M_\tot - y)$ and $\lambda_{N_0+1,0} = x$, where $x$ and $y$ remain undetermined. Assume now that all the pure state spin-densities $n^\s_\NM(\rr)$ are known; furthermore, in our specific case, the following relations hold: $n^\s_{N_0,\half}(\rr) = n^\sbar_{N_0,-\half}(\rr)$, $n^\s_{N_0+1,1}(\rr) = n^\sbar_{N_0+1,-1}(\rr)$ and $n^\s_{N_0+1,0}(\rr) = n^\sbar_{N_0+1,0}(\rr) = \thalf(n^\up_{N_0+1,1}(\rr)+n^\dw_{N_0+1,1}(\rr))$. Here $\sbar$ denotes the spin channel opposite to $\s$. The spin polarization, normally defined as $\zeta(\rr) = \frac{n_\up(\rr) - n_\dw(\rr)}{n_\up(\rr) + n_\dw(\rr)}$, equals in this case
\begin{align}
    \zeta(\rr) = \frac{\frac{y}{S_0}(n^\up_{N_0,\half}-n^\dw_{N_0,\half})+\frac{M_\tot - y}{S_1}(n^\up_{N_0+1,1}-n^\dw_{N_0+1,1})}{(1-\alpha)(n^\up_{N_0,\half}+n^\dw_{N_0,\half})+\alpha(n^\up_{N_0+1,1}+n^\dw_{N_0+1,1})}.
\end{align}
$\zeta(\rr)$ linearly depends on the parameter $y$ and is independent of $x$. In Fig.~\ref{fig:C__zeta}, $\zeta(\rr)$ is plotted for various values of~$y$, with $\alpha=0.8$ and $M_\tot=0$, using approximate densities. It is clearly seen from the plot that the ambiguity in the ensemble spin polarization is significant. Further technical details and additional plots of the polarization are provided in the Supplementary Material (SM), Sec.~I.

In this context, we wish to highlight two points. First, note that in the case of $M_\tot=0$, the polarization does not necessarily vanish; it does so only if we choose $y=0$.  Thus, zero spin does not imply a vanishing spin polarization.
Second, setting the asymptotic value of the ensemble polarization, which in
the case of the C atom equals $\zeta_\textrm{asymp} = (M_\tot -y)/(\alpha S_1)$, can serve as a way to  determine $y$; an additional criterion would be required for~$x$.

\noindent \textbf{Removing non-uniqueness. Maximal entropy.} Three ways to remove the aforementioned non-uniqueness of the ensemble ground state were discussed in Ref.~\cite{GoshenKraisler24}. For completeness, we mention them briefly below, and then suggest a new way for non-uniqueness removal: by maximization of the system's entropy.

First, it is possible to introduce additional constraints on the system, in addition to Eqs.~\eqref{eq:Tr_Lambda}, \eqref{eq:N_tot} and~\eqref{eq:M_tot}.
One way is to constrain higher powers of $\hat{S}_z$. Although mathematically this is a possible route, which eventually fully determines the ground state, the physical meaning of constraining some high powers of $\hat{S}_z$ remains obscure.

Second, it has been shown that the requirement that the deviation $\Delta S_z \equiv \sqrt{ \stat{\hat{S}_z^2} - \stat{\hat{S}_z}^2}$ is minimal, \emph{always} removes the ambiguity in the ground state, irrespectively of the number of statistical weights, i.e., of $S_0$ and $S_1$. Interestingly, for integer $N_\tot$ the resultant ground state reduces to a two-state ensemble, which, for a given $M_\tot$, are the states with the closest values of $M$. Similarly, for a fractional $N_\tot$, the ground state is comprised of three states, which correspond to three of the four vertices of the square enclosing the point $(N_\tot,M_\tot)$ (cf.~Refs.~\cite{Saavedra12,XDYang16} and see the SM, Sec.~II for a detailed derivation).

Third, in contrast to the previously mentioned approaches, introduction of an external magnetic field~$\textbf{B}$, although making the pure state energies $E_\NM$ $M$-dependent, does not always remove the non-uniqueness of the ensemble state. To remove non-uniqueness, the undetermined statistical weights have to appear in the $\textbf{B}$-dependent ensemble energy, which can then be minimized by a proper choice of these weights, thus fully determining the ground state. This is, however, not always the case, with the C atom serving as a counter-example.

In this work, we suggest a new way to uniquely define the ground state: by maximizing the electronic entropy of the system. For a system described by the ensemble $\hat{\Lambda}$ of Eq.~\eqref{eq:Lambda_def}, the von Neumann entropy equals
\begin{equation}\label{eq:S_def}
    \mathscr{S} = - \Tr{  \hat{\Lambda} \ln \hat{\Lambda} } = - \sum_{N} \sum_M \lambda_\NM \ln (\lambda_\NM).
\end{equation}

\blue{At finite temperature, one would be minimizing the free energy $F = E - T\mathscr{S}$, which in the limit $\kB T \rarr 0^+$, would mean minimizing $E_\tot$ and maximizing $\mathscr{S}$.}
We are looking therefore for such weights $\lambda_\NM$ that $\mathscr{S}$ is maximal, and conditions~\eqref{eq:lambda_0_1}, \eqref{eq:Tr_Lambda}, \eqref{eq:N_tot} and~\eqref{eq:M_tot} hold. We satisfy Condition~\eqref{eq:lambda_0_1} by setting
\begin{equation} \label{eq:lambda_NM_theta_NM}
    \lambda_\NM = \sin^2 \theta_\NM,
\end{equation}
with $0\leqs\theta_\NM\leqs \frac{\pi}{2}$, and the other three conditions via a constrained minimization with Lagrange multipliers $t_1$, $t_2$ and $t_3$:
\begin{align} \label{eq:deltaS}
    \delta \lp[ \mathscr{S} + t_1 \lp( \sum_\NM \lambda_\NM - 1 \rp) + t_2 \lp( \sum_\NM \lambda_\NM N - N_\tot \rp) +  \right. \nonumber \\
    \left. + t_3 \lp( \sum_\NM \lambda_\NM M - M_\tot  \rp)\rp] = 0
\end{align}
Substituting Eqs.~\eqref{eq:S_def} and~\eqref{eq:lambda_NM_theta_NM} into Eq.~\eqref{eq:deltaS} and taking a partial derivative with respect to each $\theta_\NM$, we reach the condition that each $N$ and $M$ must satisfy:
\begin{align}
    \sin(2 \theta_\NM) \lp[ 1 + \ln \lp( \sin^2 \theta_\NM \rp) - t_1 - t_2 N - t_3 M\rp] = 0.
\end{align}
For a given $N$ and $M$, this condition can be satisfied in three cases: (i)~If $\theta_\NM=0$, which means, however, that $\lambda_\NM=0$ (Eq.~\eqref{eq:lambda_NM_theta_NM}).
This case cannot maximize the entropy: the maximum entropy ensemble must include every allowed pure state (see the SM, Sec.~III for a proof).
(ii)~If  $\theta_\NM= \tfrac{\pi}{2}$. But in this case $\lambda_\NM=1$, which means that all the other $\lambda$'s vanish (Eq.~\eqref{eq:Tr_Lambda}). This is a pure state with entropy 0. Finally, (iii)~If we satisfy $\ln \lp( \sin^2 \theta_\NM \rp) = t_1 - 1 + t_2 N + t_3 M$, namely
\begin{align} \label{eq:maxS_reqmnt}
\lambda_\NM = e^{(t_1-1) + t_2 N + t_3 M},
\end{align}
we reach maximal entropy.

Taking into account that our ensemble $\hat{\Lambda}$ is given by Eq.~\eqref{eq:Lambda_gs}, i.e.\ the only nonzero weights are $\lambda_{N_0,-S_0}, \cdots, \lambda_{N_0,S_0}$ and $\lambda_{N_0+1,-S_1}, \cdots, \lambda_{N_0+1,S_1}$, we realize that Eq.~\eqref{eq:maxS_reqmnt} implies that
\begin{align}\label{eq:maxS_lambdas}
    \lambda_{N_0,M} = a z^M \:\:\: \textrm{and} \:\:\: \lambda_{N_0+1,M} = b  z^M,
\end{align}
where $z=e^{t_3}$, $a=e^{t_1-1+t_2 N_0}$ and $b = a e^{t_2}$.
The parameters $a$, $b$ and $z$ can be found by explicitly satisfying Eqs.~\eqref{eq:lambda_1_m_alpha}, \eqref{eq:lambda_alpha} and~\eqref{eq:lambda_Mtot}.

To show that the ground state is uniquely defined by requiring maximal entropy, we need to find $a$, $b$ and $z$ and show they are unique. To this end, we define
\begin{align}
    F(z,S) = \sum_{M=-S}^S z^M = \frac{z^{S+1/2} - z^{-(S+1/2)}}{z^{1/2} - z^{-1/2}}.
\end{align}
As a result, Eqs.~\eqref{eq:lambda_1_m_alpha} and~\eqref{eq:lambda_alpha} yield the expressions $a = (1-\alpha)/F(z,S_0)$
and $b = \alpha/F(z,S_1)$; the value of $z$ has to be determined from Eq.~\eqref{eq:lambda_Mtot}, which will then determine all three parameters.

To employ Eq.~\eqref{eq:lambda_Mtot}, we first notice that
\begin{align}
    & \sum_{M=-S}^S M z^M = z \frac{\de F}{\de z}  = \nonumber \\
    & = F(z,S) \cdot \lp[ (S+\thalf) \frac{z^{S+1/2} + z^{-(S+1/2)}}{z^{S+1/2} - z^{-(S+1/2)}} - \half \frac{z^{1/2} + z^{-1/2}}{z^{1/2} - z^{-1/2}}\rp]
\end{align}
Substituting the above relation, as well as the expressions for $a$ and $b$ into Eq.~\eqref{eq:lambda_Mtot}, we find that $z$ must satisfy the expression
\begin{widetext}
\begin{align} \label{eq:Smax_Mtot_condition}
 (1-\alpha)(S_0+\thalf) \frac{z^{S_0+1/2} + z^{-(S_0+1/2)}}{z^{S_0+1/2} - z^{-(S_0+1/2)}} + \alpha(S_1+\thalf) \frac{z^{S_1+1/2} + z^{-(S_1+1/2)}}{z^{S_1+1/2} - z^{-(S_1+1/2)}} - \half \frac{z^{1/2} + z^{-1/2}}{z^{1/2} - z^{-1/2}} = M_\tot
\end{align}
\end{widetext}
For $z \in (0, \infty)$, the left-hand side (lhs) of Eq.~\eqref{eq:Smax_Mtot_condition} is a continuous, differentiable function, and its derivative is always positive (see the SM, Sec.~IV for details). In the limit $z \rarr 0^+$, the lhs of Eq.~\eqref{eq:Smax_Mtot_condition} approaches $-M_B$ (see Eq.~\eqref{eq:MB_def}) and in the limit $z \rarr \infty$ it approaches  $M_B$. Therefore, for a given value of $M_\tot \in (-M_B, M_B)$  (i.e., within Region~\eqref{eq:spin condition}), there exists one and only one solution $z \in (0, \infty)$ to Eq.~\eqref{eq:Smax_Mtot_condition}.

As a result, $a$ and $b$ are also uniquely defined. Therefore, the requirement of maximal entropy~\eqref{eq:S_def} defines the ensemble ground state~\eqref{eq:Lambda_gs} of a many-electron system \emph{uniquely}, strictly within Region~\eqref{eq:spin condition}.
At the edges ($M_\tot=\pm M_B$), there is only one ground state anyhow, and hence no ambiguity to be removed.

The unique ensemble coefficients we have now derived, necessarily provide the global maximum of the entropy, rather than a local extremum; the entropy $\mathscr{S}$ is bounded from above, hence it must have a global maximum. By requiring that the $\lambda_\NM$ result in maximum entropy, we have arrived at a unique ensemble, which thus must be the \emph{global} maximizer of the entropy.

\begin{figure}
    \centering
\includegraphics[width=0.48\linewidth]{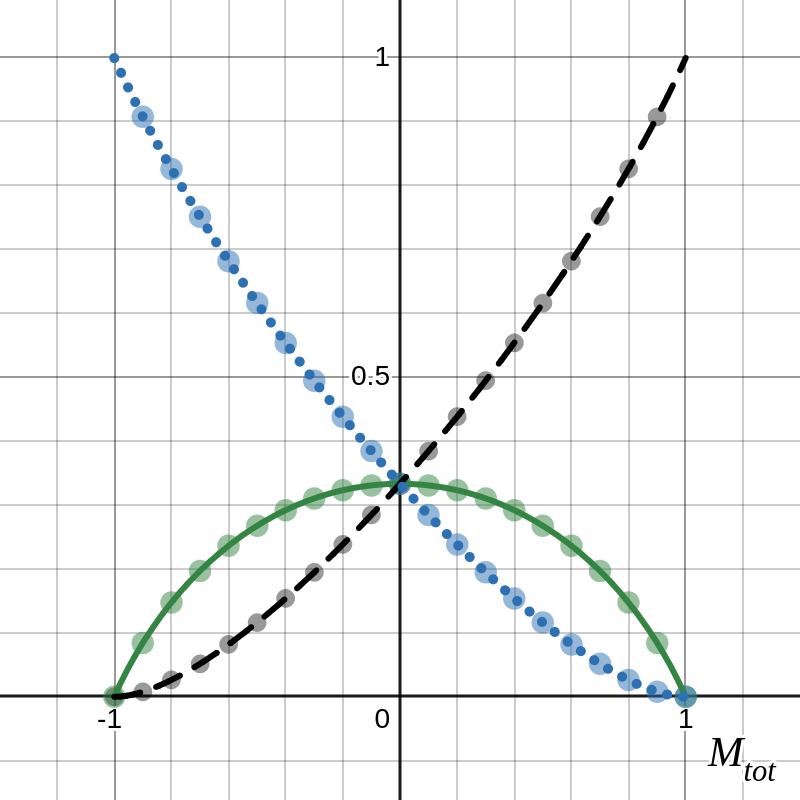}
\includegraphics[width=0.48\linewidth]{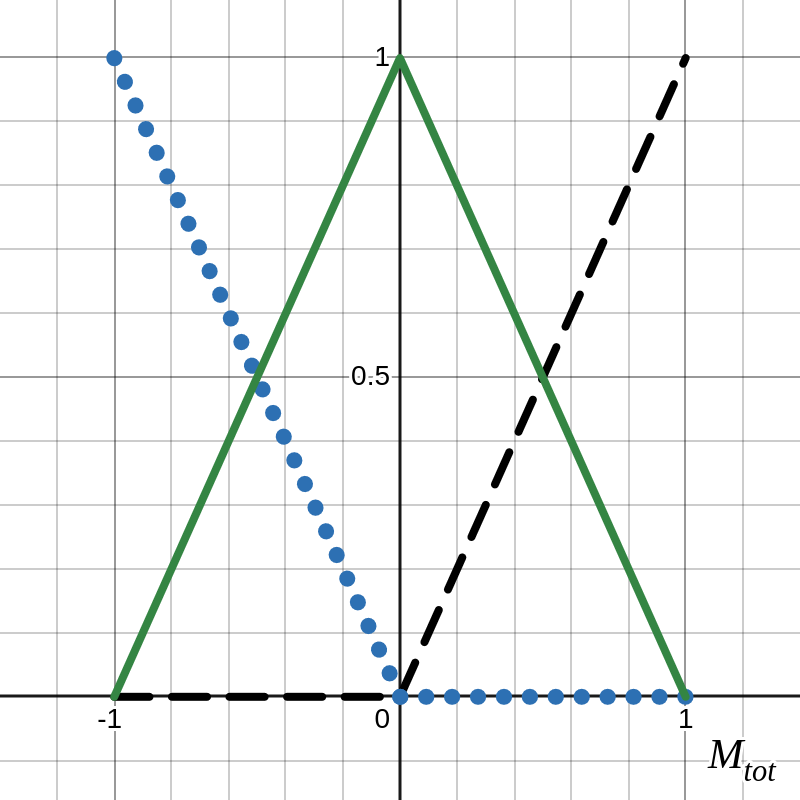}
\caption{Left: Continuous lines represent the statistical weights $\lambda_{N_0,1}$ (dashed black), $\lambda_{N_0,0}$ (solid green) and $\lambda_{N_0,-1}$ (dotted blue), versus the spin $M_\tot$, for the neutral C atom, analytically calculated from maximization of entropy for C (see Eq.~\eqref{eq:C__maxS__x}).
Circles (of the same colors) correspond to the numerical solution of Eq.~\eqref{eq:Smax_Mtot_condition}; full overlap is observed.
Right: Statistical weights $\lambda_{N_0,1}$, $\lambda_{N_0,0}$ and $\lambda_{N_0,-1}$ (same colors and dash types as on the left panel) obtained for minimal $\Delta S_z$}
    \label{fig:C__1}
\end{figure}

Let us now illustrate the consequences of the maximal entropy requirement on a specific example: the neutral C atom with varying $z$-projection of the spin, namely the case where $N_0=6$, $\alpha=0$, $S_0 = 1$ and consequently $M_\tot \in [-1,1]$.   The ensemble ground state of this system consists of three pure states, with the weights $\lambda_{N_0,1} = \thalf(1-x+M_\tot)$, $\lambda_{N_0,0} = x$ and $\lambda_{N_0,-1} = \thalf(1-x-M_\tot)$, where $x$ remains undetermined.  Maximizing the entropy either by using Eq.~\eqref{eq:Smax_Mtot_condition} or by getting an expression for the entropy for this particular case and maximizing it directly, yields
\begin{equation} \label{eq:C__maxS__x}
    x = \frac{\sqrt{4-3M_\tot^2}-1}{3}.
\end{equation}
The ensemble weights are presented graphically in Fig.~\ref{fig:C__1} (left). Particularly, for $M_\tot=0$, maximal entropy is obtained when all weights equal 1/3, as expected.  This can be shown for the general case, as well: for $M_\tot=0$, Eq.~\eqref{eq:Smax_Mtot_condition} holds for $z=1$; consequently, $F(1,S) = 2S+1$. Therefore, for the case with integer $N_\tot$, all states are equally occupied: $\lambda_{N_0,M} = 1/(2S_0+1)$, as expected. For $M_\tot=0$ and $N_\tot$ being fractional, we obtain $\lambda_{N_0,M} = (1-\alpha)/(2S_0+1)$ and $\lambda_{N_0+1,M} = \alpha/(2S_1+1)$, both being $M$-independent.

\begin{figure}
    \centering
    \includegraphics[width=0.48\linewidth]{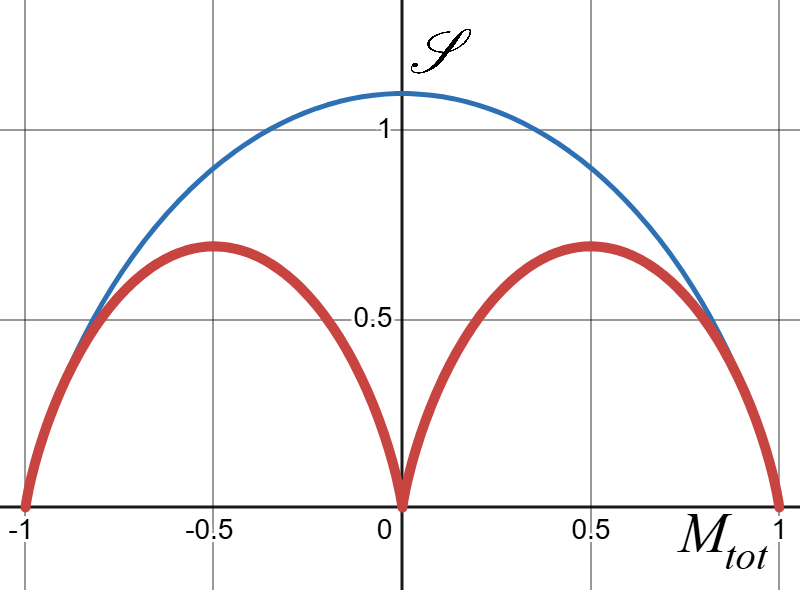}
    \includegraphics[width=0.48\linewidth]{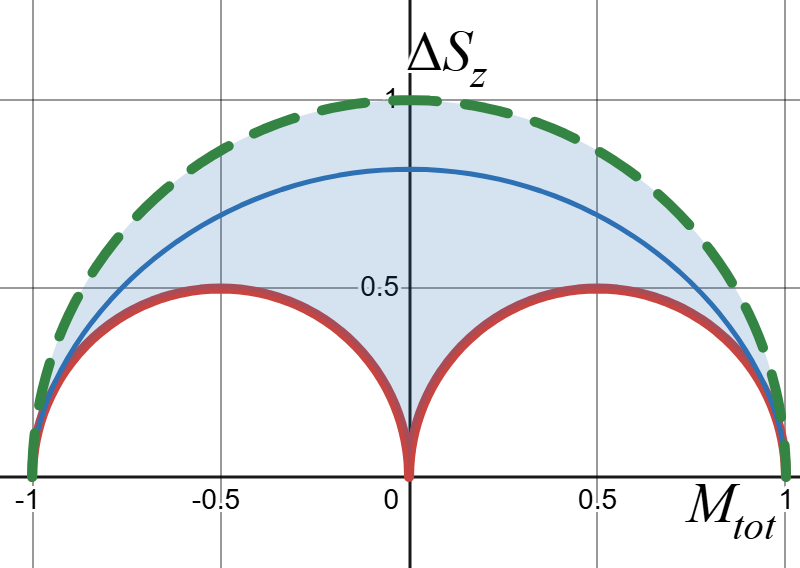}

     \caption{Left: Entropy $\mathscr{S}$ versus the spin $M_\tot$, for the neutral C atom, for the case of maximal entropy (thin blue line) and minimal $\Delta S_z$ (thick red line).
     Right: value of the deviation $\Delta S_z$ versus the spin $M_\tot$, for the case of maximal entropy (thin blue line), lying in between the lines for minimal (thick red line) and maximal (dashed green line) values of $\Delta S_z$. }
    \label{fig:C__2}
\end{figure}

We note the striking difference between  weights obtained for maximal entropy and those obtained for minimal $\Delta S_z$~\cite{GoshenKraisler24}, as shown in Fig.~\ref{fig:C__1} (right). In the maximal entropy case, the weights are nonzero everywhere, except the edges. This is in contrast to the case of only two weights being nonzero for a given value of $M_\tot$, corresponding to a gradual transition from $\ket{\Psi_{N_0,-1}}$ to $\ket{\Psi_{N_0,0}}$ and then from $\ket{\Psi_{N_0,0}}$ to $\ket{\Psi_{N_0,1}}$.
The difference between the weights in the above mentioned two scenarios is also reflected in the graphs for the entropy $\mathscr{S}$ and for the spin deviation $\Delta S_z$ (Fig.~\ref{fig:C__2}). Thus, choice of different approaches to remove ambiguity leads to substantially different results


Throughout this work, we neglect any internal degeneracy of the pure ground states at each $(N,M)$, as it does not meaningfully change our conclusions regarding the ensemble ground state.
We now emphasize that our results regarding the maximum entropy ensemble in particular, are not complicated by the addition of this degeneracy.
We discuss this scenario in detail in Sec.~V of the SM, showing that in order to maximize the entropy, the total statistical weights of each $(N,M)$ point must still be those determined by Eqs.~\eqref{eq:Smax_Mtot_condition} and \eqref{eq:maxS_lambdas}.
At each point $(N,M)$, all degenerate pure states are then given equal weights. \blue{Note that within KS-DFT, the choice of ensembles with equal weights was shown to have advantageous properties in the design of approximations~\cite{Gould_Ensemblization26}}.

\subsection{Ground state beyond Region~\eqref{eq:spin condition}} \label{sec:Beyond_Reg_1}

In this section, we reach beyond Region~\eqref{eq:spin condition} (the trapezoids), which we described so far,  to characterize the ground state also for the case of $|M_{\tot}|>M_B$. Unfortunately, beyond the trapezoids, the ensemble ground state cannot be determined exactly, but rather it depends on the specific system in question.

We focus in the following on $M_\tot > M_B$ and prove several exact statements as to the ground state there. The case $M_\tot < - M_B$ is a mirror image, and should be treated analogously.

\begin{figure}
    \centering
    \includegraphics[width=0.98\linewidth]{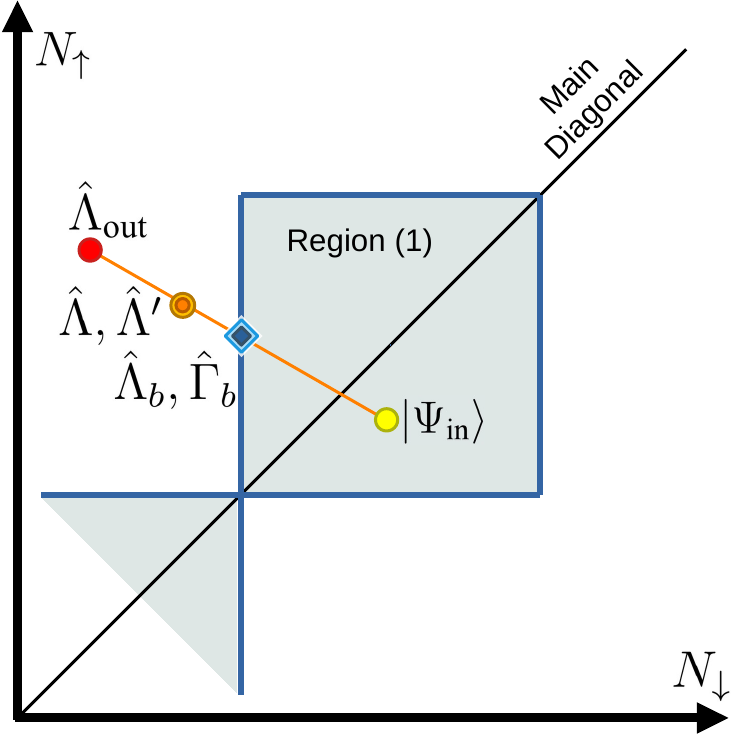}
     \caption{Illustration of the proof of Statement 1: The ensembles $\hat{\Lambda}_\out$ (red circle), $\ket{\Psi_\inn}$ (yellow circle, inside Region~\eqref{eq:spin condition}), $\hat{\Lambda}$ and $\hat{\Lambda}'$ (dark yellow and orange circles; coincide), and $\hat{\Lambda}_b$ and $\hat{\Gamma}_b$ (blue and cyan rhombuses; coincide at the edge of Region~\eqref{eq:spin condition}), depicted in the $N_\up-N_\dn$ plane.
     }
    \label{fig:stmnt1}
\end{figure}

\noindent \textbf{Statement 1.}  For $M_\tot > M_B$, the pure states $\ket{\Psi_\NM}$ that contribute to the ground-state ensemble all have $M \geqs S_{\m}(N)$. In words, when outside Region~\eqref{eq:spin condition}, use only pure states that are outside or at the edge of that region, from the correct side (to the left of the main diagonal in the $N_\up-N_\dn$ plane
~\footnote{The main diagonal in the $N_\up-N_\dn$ plane is the straight line $N_\up = N_\dw$.},
in our case). States within Region~\eqref{eq:spin condition} or from the other side of the main diagonal cannot help us.
Fig.~\ref{fig:stmnt1} presents a useful geometric illustration of the following proof.

\noindent \textbf{Proof.} Assume, aiming for a contradiction, that at least one of the pure states that contributes to the ensemble $\hat{\Lambda}$ with a nonzero weight, $\ket{\Psi_\inn} \defeq \ket{\Psi_{N_\inn,M_\inn}}$ has $M_\inn<S_{\m}(N_\inn)$.
It resides strictly to the right of the positive boundary of Region~\eqref{eq:spin condition}, i.e.\ either \textbf{in}side the trapezoids or to the right of the main diagonal. Furthermore, denote the statistical weight of this state as  $x \defeq \lambda_{N_\inn,M_\inn} \in (0,1)$. \footnote{The variables $x$ and $y$ here are unrelated to those in Sec.~\ref{sec:in_Reg_1}}

Then, we can express the ensemble as $\hat{\Lambda}= x \ketbra{\Psi_\inn} + (1-x) \hat{\Lambda}_\out$, where we defined $\hat{\Lambda}_\out$ to be the (normalized) sub-ensemble of $\hat{\Lambda}$, from which we removed $\ketbra{\Psi_{\inn}}$.
We denote the expectation values
$N_\out=\Tr \{ \hat{\Lambda}_\out \hat{N} \}$ and $M_\out=\Tr \{ \hat{\Lambda}_\out \hat{S}_z \}$, applying the usual trace operations to $\hat{\Lambda}_\out$. The latter two quantities are generally fractional.
Varying $x$ from 0 to 1 means moving the point $(N_\tot,M_\tot)$
~\footnote{Here and below it is more convenient to represent a point in the $N_\up-N_\dw$ plane by providing the values of $N_\tot$ and $M_\tot$, rather than the values of $N_\up$ and $N_\dw$. }
, which corresponds to the ensemble $\hat{\Lambda}$, on a straight line that connects the points $(N_\out,M_\out)$ ($x=0$; $\hat{\Lambda}_\out$) and  $(N_\inn, M_\inn)$ ($x=1$; $\ketbra{\Psi_{\inn}}$) (see Fig.~\ref{fig:stmnt1}).

The point $(N_\inn, M_\inn)$ is strictly within Region~\eqref{eq:spin condition}, while $(N_\tot,M_\tot)$ is strictly outside it. Therefore, the line segment connecting these points intersects the positive boundary of~\eqref{eq:spin condition} at some point $(N_b,M_b)$, i.e.\ there exists $y \in (0,1)$ such that the point $(N_b,M_b) \eqdef y (N_\inn, M_\inn) + (1-y)(N_\tot,M_\tot)$ is on the boundary of the trapezoids: $M_b=M_B(N_b)$.

We now construct an auxiliary ensemble state $\hat{\Lambda}_b \defeq y \ketbra{\Psi_\inn} + (1-y) \hat{\Lambda}$, which is instrumental to arrive at the desired contradiction. By construction, the expectation values of $\hat{\Lambda}_b$ for $\hat{N}$ and $\hat{S}_z$  are $N_b$ and $M_b$. Next, from the definition of $\hat{\Lambda}_\out$ and $\hat{\Lambda}_b$, we find, by some algebraic manipulation, that
\begin{align}\label{eq:Lambda__Lambda_b__Lambda_out}
\hat{\Lambda} = \frac{x}{y+ (1-y)x} \hat{\Lambda}_b + \frac{y(1-x)}{y+ (1-y)x} \hat{\Lambda}_\out
\end{align}

As we have seen in Eq.~\eqref{eq:statement_0}, the ground state at the boundary of the trapezoids is \emph{unique}, consisting of \emph{two} pure states which are also at the boundary. Representing the fractional $N_b$ as $N_b = N_b^0 + \alpha_b$, where $N_b^0 = \textrm{floor} (N_b)$, the ground state at the point $(N_b,M_b)$ is
\begin{align} \label{eq:Gamma_b}
    \hat{\Gamma}_b = (1-\alpha_b)  |\Psi_{N_b^0,S_b^0} \rangle \langle \Psi_{N_b^0,S_b^0}| + \alpha_b | \Psi_{N_b^0+1,S_b^1} \rangle \langle \Psi_{N_b^0+1,S_b^1}|.
\end{align}
Here $S_b^0 = S_{\m}(N_b^0)$ and $S_b^1 = S_{\m}(N_b^0+1)$.
Since $\ketbra{\Psi_\inn}$ contributes to $\hat{\Lambda}_b$ with some positive statistical weight $y$, and is not one of the pure states forming $\hat{\Gamma}_b$, we conclude that $\hat{\Lambda}_b$ \emph{cannot} be a ground state at $(N_b,M_b)$. Then,
\begin{align} \label{eq:Tr_L_b__G_b}
\Tr{\hat{\Lambda}_b \hat{H}} > \Tr{\hat{\Gamma}_b\hat{H}}.
\end{align}

Next, we construct another auxiliary ensemble state
\begin{align}
\hat{\Lambda}' = \frac{x}{y+ (1-y)x} \hat{\Gamma}_b + \frac{y(1-x)}{y+ (1-y)x} \hat{\Lambda}_\out
\end{align}
(cf.\ Eq.~\eqref{eq:Lambda__Lambda_b__Lambda_out}). The expectation values of $\hat{\Lambda}'$ for $\hat{N}$ and $\hat{S}_z$  are identical of those of $\hat{\Lambda}$. Therefore, in Fig.~\ref{fig:stmnt1}, $\hat{\Lambda}'$ and $\hat{\Lambda}$ correspond to the same point.
Then, from  Eq.~\eqref{eq:Tr_L_b__G_b} we immediately find that $\Tr{\hat{\Lambda} \hat{H}} > \Tr{\hat{\Lambda}'\hat{H}}$, in contradiction to the assumption that $\hat{\Lambda}$ is a \emph{ground} state at
$(N_\tot,M_\tot)$.

\noindent \textbf{Statement 2.} In absence of accidental degeneracy, the ensemble ground state outside the trapezoids is comprised of \emph{three} pure states, at most.

\noindent \textbf{Proof.}  The energy $\Tr{\hat{\Lambda}\hat{H}} = \sum_N \sum_M \lambda_\NM E_\NM$ is minimized under the constraints \eqref{eq:lambda_0_1}, \eqref{eq:Tr_Lambda}, \eqref{eq:N_tot} and  \eqref{eq:M_tot}, similarly to what we did in Sec.~\ref{sec:in_Reg_1} when examining the maximum entropy requirement within Region~\eqref{eq:spin condition}. Using Eq.~\eqref{eq:lambda_NM_theta_NM} for the coefficients $\lambda_\NM$ and performing a constrained minimization with $z_1$, $z_2$ and $z_3$ as Lagrange multipliers,
we obtain
\begin{align}
&\delta \left[ \sum_\NM \lambda_\NM E_\NM + z_1 \lp( \sum_\NM \lambda_\NM
 -1 \rp) \right. \nonumber \\
& \left. +z_2 \lp( \sum_\NM  \lambda_\NM N - N_\tot \rp)+ z_3 \lp( \sum_\NM  \lambda_\NM M - M_\tot \rp) \rp]=0.
\end{align}
Taking the derivative with respect to each $\theta_\NM$, we get
\begin{equation}
 \sin(2\theta_\NM) \lp[ E_\NM + z_1  +z_2 N +z_3 M \rp] =0
\end{equation}
Again, there are three ways to satisfy each such constraint: First, $\lambda_{\NM}=0$, i.e., this particular pure state does not contribute to the ensemble. Second, $\lambda_{\NM}=1$, which means that all the other ensemble coefficients are zero, i.e., we are actually dealing with a pure state, and Statement 2 trivially holds. Third, we must satisfy
\begin{equation}\label{eq:Lagrange_mult}
    E_\NM + z_1  + z_2 N + z_3 M =0,
\end{equation}
for all relevant $N$ and $M$. If there are $k$ pure states $\ket{\Psi_1},\ldots,\ket{\Psi_k}$ that are contributing to the ensemble, then Eqs.~\eqref{eq:Lagrange_mult} are $k$ linear equations for the three Lagrange multipliers. If $k>3$, there is generally \emph{no solution} for this set of equations. It therefore follows that the ensemble $\hat{\Lambda}$ includes three pure states, at most.

Note that outside the trapezoids, states $\ket{\Psi_{\NM}}$ with $M<0$ cannot participate in the solution (see Statement 1). This is in contrast to the situation within the trapezoids, where states with $M>0$ and $M<0$
can both participate in the ensemble due to the equality between their energies, $E_{N,M}=E_{N,-M}$. Within the trapezoids, Eqs.~\eqref{eq:Lagrange_mult} can be satisfied for more than three pure state ingredients, with $z_3=0$.

Notably, there may occur the improbable situation of \emph{accidental degeneracy} outside the trapezoids, where more than three pure states do contribute to the ensemble. This is the situation where the energies $E_\NM$ satisfy certain relations between them, such that Eqs.~\eqref{eq:Lagrange_mult} automatically hold for more that three pure states.
Namely, this occurs if the points $(N,M,E_\NM)$ lie on the same plane, for more than three relevant values of $N$ and $M$. A particular example for such a relation between $E_\NM$'s is given below, in Sec.~\ref{sec:CasesPsiStar}.

Another feature of the ground state outside the trapezoids is the fact that there is no ambiguity in the ground state, unlike the situation within Region~\eqref{eq:spin condition} (see Sec.~\ref{sec:in_Reg_1}). This is a consequence of Statement~2: the number of states is not higher than the number of Constraints \eqref{eq:Tr_Lambda}, \eqref{eq:N_tot} and~\eqref{eq:M_tot}.

Finally, note that the three states that comprise the ensemble form a triangle in the $N_\up - N_\dn$ plane. To satisfy Constraints \eqref{eq:lambda_0_1}, \eqref{eq:Tr_Lambda}, \eqref{eq:N_tot} and~\eqref{eq:M_tot}, this triangle must include the point $(N_\tot,M_\tot)$. This geometric observation is useful to determine which pure states are relevant for the ensemble, in specific cases. Particularly, we found it useful to remember that if one draws a vertical line that passes through the point $(N_\tot, M_\tot)$, the three points that correspond to the three pure states of the ensemble cannot land on one side of such a line, because then such an ensemble can never yield the correct $N_\dw$. Similar condition is true if drawing a horizontal line, a line parallel to the main diagonal and a line perpendicular to the main diagonal, through $(N_\tot, M_\tot)$.

As a consequence of Statement~2, we can express the ground state as
\begin{equation} \label{eq:Lambda_123}
\hat{\Lambda}=\lambda_1 \ketbra{\Psi_1}+\lambda_2\ketbra{\Psi_2}+\lambda_3\ketbra{\Psi_3},
\end{equation}
being a sum of three, generally unknown, pure ground states that reside on the lattice of the $N_\up-N_\dn$ plane, in the region described in Statement 1. As a result, the expectation value of any operator $\hat{O}$ is $O_\ens = \Tr{\hat{\Lambda}\hat{O}}=\lambda_1 O_1 +\lambda_2 O_2 +\lambda_3 O_3$, where $O_i:=\expval{\hat{O}}{\Psi_i}$ ($i = 1, 2, 3$).
Satisfying Constraints \eqref{eq:Tr_Lambda}, \eqref{eq:N_tot} and \eqref{eq:M_tot}, which now can be expressed as $\lambda_1 + \lambda_2 + \lambda_3 = 1$, $\lambda_1 N_1 + \lambda_2 N_2 + \lambda_3 N_3 = N_\tot$ and $\lambda_1 M_1 + \lambda_2 M_2 + \lambda_3 M_3 = M_\tot$, we find that the ensemble expectation value $O_\ens$ equals
\begin{align} \label{eq:O_ens}
  & O_\ens =  \frac{1}{D} \cdot \nonumber \\
   \cdot \Big\{ & \lp[ (N_\tot-N_3)(M_2-M_3) - (N_2-N_3)(M_\tot-M_3)\rp] O_1 \nonumber \\
   + & \lp[ (N_\tot-N_1)(M_3-M_1) - (N_3-N_1)(M_\tot-M_1)\rp] O_2  \nonumber \\
    + &  \lp[ (N_\tot-N_2)(M_1-M_2) - (N_1-N_2)(M_\tot-M_2)\rp] O_3 \Big\},
\end{align}
where $D = M_1(N_3-N_2) + M_2(N_1-N_3) + M_3(N_2-N_1)$.
Therefore, outside the trapezoids, the profile of any ensemble quantity, including the energy, is piecewise-linear in $N_\tot$ and $M_\tot$: it is a collection of tiles -- segments of planes -- whose boundaries are determined, for each particular system, by the relationships between its pure state total energies, $E_\NM$, while satisfying Constraints \eqref{eq:Tr_Lambda}, \eqref{eq:N_tot} and \eqref{eq:M_tot}.
Within each such tile, $O_\ens$ is a linear function of in $N_\tot$ and $M_\tot$.
Contrary to the situation within the trapezoids, here the condition $ [ \hat{O}, \hat{S}_+ ] = 0 $ is not required.

So far, we described the general case outside the trapezoids. In the following, we focus on the particular, yet important case, where the point $(N_\tot, M_\tot)$ is immediately outside Region \eqref{eq:spin condition}, and $N_\tot$ is strictly fractional. This means that if we express the ensemble spin as $M_\tot=M_B+\delta$, then $\delta>0$ is arbitrarily small. The importance of this case is, among other things, in the description of spin-migration derivative discontinuities~\cite{GoshenKraisler24,Hayman25}.

\noindent \textbf{Statement 3.} For an ensemble state at $(N_\tot, M_\tot)$, with $N_\tot=N_0+\alpha$ and $\alpha \in (0,1)$, and ${M_\tot=M_B+\delta}$, where $\delta$ is sufficiently small,  two of the pure states involved in the ensemble are $\ket{\Psi_{N_0,S_0}}$ and $\ket{\Psi_{N_0+1,S_1}}$. Furthermore, the identity of the third pure state involved does not depend on $\alpha$.

\noindent \textbf{Proof.}
At $M_\tot=M_B$, we already know from Eq.~\eqref{eq:statement_0} that the only pure states that contribute to the ensemble with nonzero coefficients are $\ket{\Psi_{N_0,S_0}}$ and $\ket{\Psi_{N_0+1,S_1}}$.
The ensemble coefficients $\lambda_\NM$ are continuous functions of $N_\tot$ and $M_\tot$
~\footnote{Otherwise, the ensemble energy and other physical quantities, which are linear functions of $\lambda_\NM$, would become almost inevitably discontinuous over the $N_\up-N_\dn$ plane.},
and therefore, for sufficiently small values of $\delta$, $\lambda_{N_0,S_0}$ and $\lambda_{N_0+1,S_1}$ will remain finite.
This means that we identified two of the three states of Eq.~\eqref{eq:Lambda_123}, and the additional pure state, $\ket{\Psi_\star} \defeq \ket{\Psi_{N_\star,M_\star}}$, is to be determined by minimization of the energy, scanning all the possible pure states.

Hence, in the case of small $\delta$, using Eq.~\eqref{eq:O_ens} and substituting $N_1=N_0$, $N_2=N_0+1$, $M_1=S_0$ and $M_2=S_1$, $N_\tot = N_0+\alpha$ and $M_\tot = M_B + \delta$, the ensemble energy takes the form
\begin{equation} \label{eq:E_delta_star}
    E_{\tot}=(1-\alpha)E_{N_0}+\alpha E_{N_0+1}+ \delta \, \Delta E_\star,
\end{equation}
where
\begin{align} \label{eq:DE_star}
&\Delta E_\star = \frac{(N_\star - N_0) I +(E_\star-E_{N_0})} {(N_\star - N_0)(S_0-S_1) + (M_\star-S_0)}
\end{align}
and $I = E_{N_0}-E_{N_0+1}$ is the IP of the $(N_0+1)$-system. [It is equivalent to $A_1$ -- the affinity of the $N_0$-system, as denoted above.]

The energy in Eq.~\eqref{eq:E_delta_star} has to be minimized with respect to the choice of the pure state $\ket{\Psi_\star}$ (or equivalently $N_\star$ and $M_\star$), meaning that one has to minimize just $\Delta E_\star$. Notably, this term does not explicitly depend on $\alpha$. Furthermore, the set of all allowed pairs ($N_\star, M_\star)$ does not depend on~$\alpha$ either, as we now show:
The allowed states $(N_\star, M_\star)$ are those for which $\lambda_i\in[0,1]$. In the case of small $\delta$, the only remaining requirement which is not identically satisfied,
is that $\lambda_3 = \delta / ((N_\star-N_0)(S_0-S_1)+(M_\star-S_0))$ is positive. It can be expressed as
\begin{equation}
M_\star > (N_0+1 - N_\star) S_0 + (N_\star-N_0)S_1.
\label{eq:psi_star_ineq}
\end{equation}
This condition does not depend on the choice of $\alpha$. Then, once the state $\ket{\Psi_\star}$ is found for a given $\alpha$, it is the same for all other $\alpha$'s, as well.
Our proof is therefore concluded: the third state that minimizes the ensemble energy (and keeps $\lambda_i\in[0,1]$) will be the same for all values of $\alpha$.

The allowed $\ket{\Psi_\star}$'s must satisfy Eq.~\eqref{eq:psi_star_ineq}, as we discussed above, but also they must be outside the trapezoids, by Statement 1. These are two independent constraints: Eq.~\eqref{eq:psi_star_ineq} means $\ket{\Psi_\star}$ is on the positive side (in terms of $M$) of the line connecting the points $(N_0,S_0)$ and $(N_0+1,S_1)$ in the $N_\up-N_\dw$ plane, whereas the constraint of $\ket{\Psi_\star}$ residing outside the trapezoids, does not imply, and is not implied by Eq.~\eqref{eq:psi_star_ineq}.
The energy in Eq.~\eqref{eq:DE_star} is to be minimized over the set of all $\ket{\Psi_\star}$'s  which satisfy both these constraints.

Although we have discussed here the limit where $\delta$ is arbitrarily small, in practice Statement 3 holds for a finite range of $\delta$ values. Namely, as long as the point $(N_\tot,M_\tot)$ is inside the triangle whose vertices are $(N_0,S_0), (N_0+1,S_1)$ and $(N_\star,M_\star)$, the ensemble is composed of the same three pure states corresponding to these points. In this way, the range of $\delta$ values where Statement 3 holds for a given system depends on $\alpha$ and $(N_\star,M_\star)$.

\section{Ensemble states beyond Region~(\ref{eq:spin condition}): Particular cases of~$\ket{\Psi_\star}$} \label{sec:CasesPsiStar}

\begin{figure}
    \centering
    \includegraphics[width=0.494\linewidth]{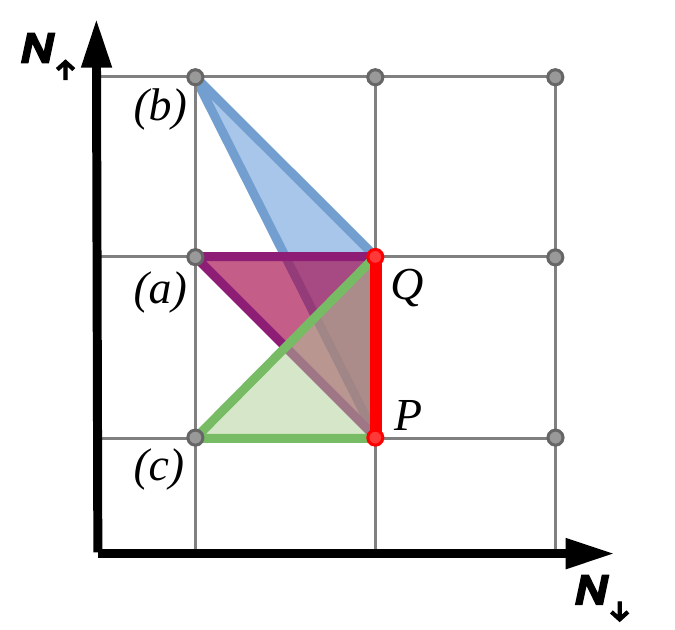}
\includegraphics[width=0.494\linewidth]{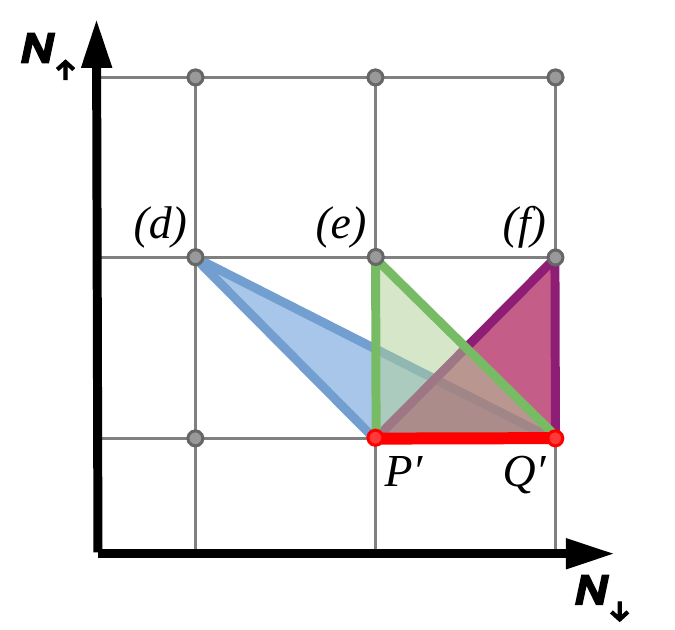}
    \caption{Illustration of possible particular cases for the state $(N_\star,M_\star)$ (defined in text) and resultant triangular tiles in the $N_\up - N_\dw$ plane, for $S_1 = S_0+ \half$ (left panel) and $S_1 = S_0 - \half$ (right panel).     }
    \label{fig:commonThreePsiStar}
\end{figure}

We have formally shown in Sec.~\ref{sec:Beyond_Reg_1} that in the limit $M_\tot\to M_B^+$, two of the three ensemble states are known, and the third one, $\ket{\Psi_\star}$, has to be determined out of all possible states, such that the ensemble energy of Eq.~\eqref{eq:E_delta_star} is minimized.
Now, we illustrate the situation beyond the trapezoids, in the limit $M_\tot\to M_B^+$, by considering common cases of $\ket{\Psi_\star}$'s. We address two scenarios: adding an $\up$-electron to a $N_0$-system, i.e.\ the scenario when the equilibrium spin of the $(N_0+1)$-system is higher by half than that of the $N_0$-system ($S_1 = S_0 + \half$) and adding a $\dw$-electron, i.e.\ when the equilibrium spin of the $(N_0+1)$-system is lower by half ($S_1 = S_0 - \half$). A graphical illustration is provided in Fig.~\ref{fig:commonThreePsiStar}.

For $S_1 = S_0 + \half$, the two known states, denoted by points $P = (N_0, S_0)$ and $Q = (N_0+1,S_1)$ are connected in Fig.~\ref{fig:commonThreePsiStar} (left panel) by a red vertical line. Three common cases for the state $\ket{\Psi_\star}$ are considered: (a) $(N_0,S_0+1)$, in purple. Its total energy is $E_{N_0}+J_{1,N_0}$; here and below $J_{k,N}=E_{N,S_{\textrm{min}}(N)+k} - E_{N,S_{\textrm{min}}(N)+k-1}$ is the $k^\textrm{th}$ spin-flip energy of a system with $N$ electrons.
(b) $(N_0+1,S_1+1)$, in blue. Its energy is $E_{N_0+1}+J_{1,N_0+1}$.
(c) $(N_0-1,S_1)$, in green.  If the equilibrium spin of the $(N_0-1)$-system is $S_{-1}=S_{\textrm{min}}(N_0-1) = S_0+\half \equiv S_1$, the energy of this state is simply $E_{N_0-1}$. If, however, $S_{-1} = S_0-\half$, then the energy of this state is $E_{N_0-1}+J_{1,N_0-1}$.  If we consider additional states, e.g., $(N_0-2,S_0)$, the energies for them can be derived in a similar manner.

To compare the total energies of Cases (a), (b) and (c), it is enough to compare the corresponding $\Delta E_\star$ of Eq.~\eqref{eq:DE_star}, which equal $\Delta E_{\textrm{(a)}} = J_{1,N_0}$, $\Delta E_{\textrm{(b)}} = J_{1,N_0+1}$ and $\Delta E_{\textrm{(c)}} = E_g$ if $S_{-1}=S_0+\half$ and $\Delta E_{\textrm{(c)}} = E_g+J_{1,N_0-1}$ if $S_{-1}=S_0-\half$. Here $E_g = E_{N_0+1} - 2E_{N_0} + E_{N_0-1}$ is the fundamental gap of the $N_0$-system.  Therefore, it is the relation between the fundamental gap and the spin-flip energies of the $(N_0-1)$-, $N_0$- and $(N_0+1)$-systems that defines which $\ket{\Psi_\star}$ prevails. Since to the best of our knowledge there is no clear relation between the aforementioned quantities (unlike e.g.\ the convexity conjecture for IPs), various ensembles outside the trapezoids are possible (see Sec.~\ref{sec:NumInvestigation} below).

In the scenario of $S_1 = S_0 - \half$, the situation is similar. The two known states,  $P' = (N_0, S_0)$ and $Q' = (N_0+1,S_1)$, are connected in Fig.~\ref{fig:commonThreePsiStar} (right panel) by a red horizontal line. For the states (d) $(N_0, S_0+1)$, (e) $(N_0+1,S_1+1)$ and (f) $(N_0+2,S_0)$, which create triangular tiles
depicted in blue, green and purple, respectively, one has to compare $J_{1,N_0}$, $J_{1,N_0+1}$ and either $\tilde{E}_g$ or $\tilde{E}_g+J_{1,N_0+2}$ (depending on $S_2$). Here $\tilde{E}_g$ is the fundamental gap of the $(N_0+1)$-system.

Lastly, let us discuss the improbable scenario of \emph{accidental degeneracy} mentioned in Sec.~\ref{sec:Beyond_Reg_1}.
This scenario occurs precisely if two (or more) of the $\Delta E_\star$'s, corresponding to two different $\ket{\Psi_\star}$'s, equal each other. Then, these $\ket{\Psi_\star}$'s must both participate in the ensemble. For example, for a system in Case (a), the state $(N_0+1, S_1+1)$ joins the ensemble as a fourth state if $\Delta E_{\textrm{(b)}} = \Delta E_{\textrm{(a)}}$, which implies $J_{1,N_0} = J_{1,N_0+1}$.   Similarly, the condition for the state $(N_0, S_0+2)$ to join is $J_{2,N_0} = J_{1,N_0}$, and for the state $(N_0-1,S_{-1})$ is $E_g = J_{1,N_0}(S_1 - 2S_0 + S_{-1})$.

The above examples are special cases of Statement~2, with more than three pure states. Equation~\eqref{eq:Lagrange_mult} holds for all the participating $(N,M)$ states exactly, due to the special relations between the equilibrium spins, spin-flip energies and fundamental gaps.
Although such or similar relations do not usually hold, one should keep in mind that if this becomes the case (accidentally or by design, e.g., in some particular model system), additional state(s) must join, and the ensemble beyond Region~\eqref{eq:spin condition} will consist of more than three pure states.

\section{KS eigenvalues, potentials and discontinuities} \label{sec:eigenvalues}

Up to this point, we have derived general properties of many-electron systems. We now discuss the implications of Statements 1, 2 and 3 of Sec.~\ref{sec:Beyond_Reg_1} for quantities in KS-DFT, particularly the KS eigenvalues $\eps_{i}^{\s}$ and the KS potentials, $v_\KS^\s(\rr)$.

As to the eigenvalues, from Janak's theorem~\cite{Janak78}, which identifies $\eps_{i}^{\s}$ with derivatives of the total energy with respect to the corresponding occupation numbers, $f_{i}^{\s}$, the highest (partially) occupied $\up$ and $\dn$ KS orbital energies equal
\begin{equation}
\label{eq:eps_ho_dE_dN}
    \eps_\ho^\s = \frac{\de E}{\de f_\ho^\s} = \lp( \frac{\de E}{\de N_\s} \rp)_{N_\sbar},
\end{equation}
assuming that a change in $N_\s$ results in a change of occupation in only one KS level. Using the chain rule, we find that
\begin{align}
\lp( \frac{\de E}{\de N_\s} \rp)_{N_\sbar} = \lp( \frac{\de E}{\de N_\tot} \rp)_{M_\tot} \lp( \frac{\de N_\tot}{\de N_\s} \rp)_{N_\sbar} + \nonumber \\
+ \lp( \frac{\de E}{\de M_\tot} \rp)_{N_\tot} \lp( \frac{\de M_\tot}{\de N_\s} \rp)_{N_\sbar}.
\end{align}
We define $d_\s$ to equal 1 for $\s=\up$ and $-1$ for $\s=\dw$, and arrive at the relation
\begin{equation} \label{eq:e_ho_s}
    \eps_\ho^\s = \lp( \frac{\de E}{\de N_\tot} \rp)_{M_\tot} + \frac{d_\s}{2} \lp(\frac{\de E}{\de M_\tot} \rp)_{N_\tot}.
\end{equation}
Moving to the ``center-of-mass system'' for the ho-levels, we note that $\half (\eps_\ho^\up + \eps_\ho^\dw) = \lp( {\de E}/{\de N_\tot} \rp)_{M_\tot}$ and $\eps_\ho^\up - \eps_\ho^\dw = \lp({\de E}/{\de M_\tot} \rp)_{N_\tot}$.

Strictly within $|M_{\tot}| < M_B$, the total energy is piecewise-linear in $N_\tot$ and does not depend on $M_\tot$ (Eq.~\eqref{eq:E_tot_PL}). As a direct consequence, we found in Ref.~\cite{GoshenKraisler24} that both ho energy levels coincide and equal the negative of the IP of the $(N_0+1)$-system:
\begin{equation} \label{eq:e_ho_s_I}
\eps_{\ho}^{\up}=\eps_{\ho}^{\dn} = - I.
\end{equation}

Outside the trapezoids, $\eps_\ho^\s$ are not equal anymore, but they are constant, with respect to $N_\tot$ and $M_\tot$, within each tile: Differentiating Eq.~\eqref{eq:O_ens}, with $O$ taken to be the energy, Eq.~\eqref{eq:e_ho_s} becomes
\begin{align}\label{eq:eps_ho_123}
    \nonumber
    \eps_\ho^\s = \frac{1}{D} \bigg[ (M_2-M_3)E_1 + (M_3-M_1)E_2 +(M_1-M_2)E_3 -
    \\
    -\frac{d_\s}{2} \lp( (N_2-N_3)E_1 +(N_3-N_1)E_2 +(N_1-N_2)E_3 \rp) \bigg],
\end{align}
being independent of $N_\tot$ and $M_\tot$.  Expressions for $\half (\eps_\ho^\up + \eps_\ho^\dw)$ and $\eps_\ho^\up - \eps_\ho^\dw$ readily follow.

\begin{figure}
    \centering
    \includegraphics[width=0.98\linewidth]{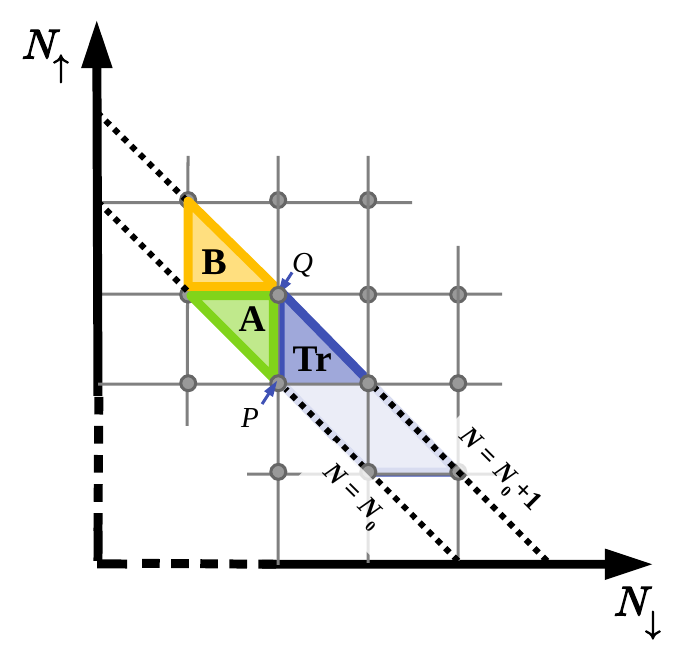}
    \caption{Illustration of Tile A (green), Tile B (yellow) and the Trapezoid region (denoted Tr; blue), for the particular case where $S_1=S_0+\half$, as discussed in text, in the $N_\up-N_\dw$ plane. The diagonals $N=N_0$ and $N=N_0+1$ and the points $P = (N_0, S_0)$ and $Q = (N_0+1,S_1)$ are explicitly presented.}
    \label{fig:Tiles_A_B}
\end{figure}

Let us now focus on the case depicted in Fig.~\ref{fig:Tiles_A_B}. Here, we set $S_1 = S_0 + \half$; the left boundary of the trapezoid (denoted Tr in Fig.~\ref{fig:Tiles_A_B}) that connects the points $P = (N_0, S_0)$ and $Q = (N_0+1,S_1)$ is a vertical line in the $N_\up-N_\dw$ plane. We further assume that
the ensemble state in the region immediately to the left of the boundary $PQ$ (denoted Tile A) is given by $\ket{\Psi_{N_0,S_0}}$, $\ket{\Psi_{N_0+1,S_1}}$ and $\ket{\Psi_{N_0,S_0+1}}$. In addition, we assume that the ensemble state in the region denoted Tile~B is given by $\ket{\Psi_{N_0+1,S_1}}$, $\ket{\Psi_{N_0,S_0+1}}$ and $\ket{\Psi_{N_0+1,S_1+1}}$. This is a particular, but a rather common case, e.g., this is the situation in Li with $2+\alpha$ electrons or in C with $5+\alpha$ electrons.
In this case, strictly within Tile A, by substituting the values of the energy, spin and electron number of the three constituent pure states within Tile A into Eq.~\eqref{eq:eps_ho_123}, we find that
\begin{equation} \label{eq:e_ho_s_IJ_Tile_A}
\eps_\ho^\up = -I \:\:\: ; \:\:\: \eps_\ho^\dw = -I - J_{1,N_0}
\end{equation}
Strictly within Tile B, by an analogous substitution into Eq.~\eqref{eq:eps_ho_123}, we find that
\begin{equation} \label{eq:e_ho_s_IJ_Tile_B}
\eps_\ho^\up = -I - J_{1,N_0} + J_{1,N_0+1} \:\:\:  ;  \:\:\: \eps_\ho^\dw = -I - J_{1,N_0}.
\end{equation}
Equations \eqref{eq:e_ho_s_I}, \eqref{eq:e_ho_s_IJ_Tile_A} and \eqref{eq:e_ho_s_IJ_Tile_B} serve as exact conditions for the highest occupied KS levels, within the case we describe (Fig.~\ref{fig:Tiles_A_B}). They can be viewed as a generalization of the IP theorem~\cite{PPLB82,PerdewLevy83,LevyPerdewSahni84,AlmbladhVonBarth85,NATO85_Perdew,PerdewLevy97,Harbola99,HodgsonKraisler17,KraislerHodgson21} to systems with an arbitrary, possibly fractional,  electron number and spin,
connecting KS eigenvalues and total energy differences. Approximate DFT calculations should aim to reconstruct these relations.

We note in passing that in the complimentary case $S_1 = S_0 - \half$, where the upper boundary of the trapezoid that connects the points $P' = (N_0, S_0)$ and $Q' = (N_0+1,S_1)$ is a horizontal line in the $N_\up-N_\dw$ plane, and the ensemble state in the region immediately above $P'Q'$ (Tile $A'$) is given by $\ket{\Psi_{N_0,S_0}}$, $\ket{\Psi_{N_0+1,S_1}}$ and $\ket{\Psi_{N_0+1,S_1+1}}$, and the ensemble state in Tile~$B'$ is given by $\ket{\Psi_{N_0,S_0}}$, $\ket{\Psi_{N_0+1,S_1+1}}$ and $\ket{\Psi_{N_0,S_0+1}}$, we obtain analogous relations for the orbitals energies. For Tile~$A'$,
\begin{equation} \label{eq:e_ho_s_IJ_Tile_A_prime}
\eps_\ho^\up = -I + J_{1,N_0+1} \:\:\: ; \:\:\: \eps_\ho^\dw = -I;
\end{equation}
For Tile~$B'$,
\begin{equation} \label{eq:e_ho_s_IJ_Tile_B_prime}
\eps_\ho^\up = -I + J_{1,N_0+1} \:\:\:  ;  \:\:\: \eps_\ho^\dw = -I +J_{1,N_0+1} - J_{1,N_0}.
\end{equation}

Due to the similarity of the above two cases, in the following we focus only on the case $S_1=S_0+\half$, depicted in Fig.~\ref{fig:Tiles_A_B}. In particular, if $\alpha \in (0, 1)$, as we cross from the trapezoid (Tr) to Tile~A, and from Tile~A to Tile~B, the identity of the ho-$\s$ levels changes. In the transition Tr $\rarr$ A, this occurs only for the ho-$\dn$ level, as $N_\dn$ passes through an integer; in the transition A~$\rarr$~B, this occurs only to the ho-$\up$ level. For example, for the Li atom with $2+\alpha$ electrons, in the trapezoid we are moving electrons from the 2$s^\dw$ KS level to 2$s^\up$. Therefore, ho-$\up$ is 2$s^\up$ and  ho-$\dw$ is 2$s^\dw$. However, in Tile A, ho-$\up$ is still 2$s^\up$, whereas ho-$\dw$ is now 1$s^\dw$. Finally, in Tile B ho-$\up$ is 2$p^\up$ and  ho-$\dw$ is still 1$s^\dw$.

Furthermore, $\eps_\ho^\s$ change abruptly, due to the discontinuity in the slope of the energy profile and the necessity to satisfy Conditions~\eqref{eq:e_ho_s_I}, \eqref{eq:e_ho_s_IJ_Tile_A} and \eqref{eq:e_ho_s_IJ_Tile_B}.  Taking the Li atom as an example again, within the trapezoid the constraint on the KS eigenvalues is $\eps_{2s}^\up = \eps_{2s}^\dw = -I$; here $I$ is the first IP of neutral Li. Within Tile A, though, there is a different constraint: $\eps_{2s}^\up = - I$ and $\eps_{1s}^\dw = -I - J_{1,2}$, where $J_{1,2}$ is the first spin-flip energy of Li$^+$.
Satisfying the former constraint does not imply the latter. Within Tile~B, we have yet another constraint: $\eps_{2p}^\up = - I - J_{1,2} + J_{1,3}$ and $\eps_{1s}^\dw = -I - J_{1,2}$, where $J_{1,3}$ is the first spin-flip energy of Li. We clearly see, therefore, that the KS energy levels need to jump.

An abrupt change in the value of a KS level, $\eps_i^\s$, necessitates a jump in the corresponding KS potential, $v_\KS^\s (\rr)$. In a similar fashion to the famous derivative discontinuity of Ref.~\cite{PPLB82}, which emerges as the electron number surpasses an integer~\cite{PPLB82, ShamSchluter83, PerdewLevy83, Perdew85, ZhangYang00, LeinKummel05, Mundt05, Cohen12, Baerends13, KraislerKronik13, GouldToulouse14, MoriS14, Mosquera14, Mosquera14a, KraislerKronik14, Goerling15}, a discontinuous change in the KS potentials must be a spatial constant, $\Delta^\s$. Otherwise, the potentials on either side of the boundary where the discontinuity occurs, which yield the same spin-densities $n_\s(\rr)$, would violate the Hohenberg-Kohn theorem~\cite{HK64}. Moving further away from the boundary, the spatially uniform jump evolves into a plateau~\cite{GouldToulouse14, HodgsonKraisler17, KraislerSchild20, KraislerHodgson21, KocakKraislerSchild21JPCL, KocakKraislerSchild23, RahatKraisler25}).

Considering the potential on both sides of any boundary between two general tiles $U$ and $V$, $v_{\KS,U}^\s (\rr)$ and $v_{\KS,V}^\s (\rr)$, respectively, we identify the discontinuity as
\begin{align}\label{eq:def_Delta}
\Delta_{U \rarr V}^\s = v_{\KS,V}^\s(\rr) - v_{\KS,U}^\s(\rr)
= (\eps_i^\s)_V - (\eps_i^\s)_U.
\end{align}
We emphasize that the above eigenvalues and KS potentials are evaluated within tiles $U$ and $V$ respectively, at adjacent points on either side of the boundary. We further note that $\Delta^\s_{U\to V}$ is not constant \emph{along} the boundary, but depends on the specific point at which we cross it.

In the case of Li discussed above, in the transition Tr~$\rarr$~A, we find that $\Delta_{\textrm{Tr} \rarr \textrm{A}}^\up = 0$ and $\Delta_{\textrm{Tr} \rarr \textrm{A}}^\dw = (\eps_{1s}^\dw)_\textrm{A} - (\eps_{1s}^\dw)_\textrm{Tr} = - (J_{1,2} - J_{1,2}^\KS)$, where $J_{1,2}^\KS \defeq (\eps_{2s}^\up - \eps_{1s}^\dw)_\textrm{Tr}$ is the KS analog of the first spin-flip energy of Li$^+$. Indeed, thinking of a non-interacting system, the spin-flip is achieved by moving an electron from the $\dw$-homo to the $\up$-lumo, which is exactly what $J_{1,2}^\KS$ represents for Li$^+$.

In the transition A~$\rarr$~B, we find that $\Delta_{\textrm{A} \rarr \textrm{B}}^\dw = 0$, while $\Delta_{\textrm{A} \rarr \textrm{B}}^\up = (\eps_{2p}^\up)_\textrm{B} - (\eps_{2p}^\up)_\textrm{A} = J_{1,3} - J_{1,2} - (J_{1,3}^\KS - J_{1,2}^\KS)$, where \blue{$J_{1,3}^\KS \defeq (\eps_{2p}^\up - \eps_{1s}^\dw)_\textrm{Tr}$} is the KS analog of the first spin-flip energy of neutral Li. We notice the similarity in the expressions for $\Delta^\s$ derived here to the well-known derivative discontinuity $\Delta = E_g - (\eps_\lu - \eps_\ho)$, where $\Delta$ is the difference between the fundamental gap of the interacting system and its KS analog.

So far we considered the transitions Tr$\to$A and A$\to$B, where $N_\tot$ is strictly fractional. For integer $N_\tot$ (in our example, this is $N_\tot=3$), when varying $M_\tot$, we encounter a single transition Tr$\to$B, instead. In this case, both potentials jump, and we find that $\Delta^\s_{\textrm{Tr} \rarr \textrm{B}} = \Delta^\s_{\textrm{Tr} \rarr \textrm{A}} + \Delta^\s_{\textrm{A} \rarr \textrm{B}}$ (cf.~Ref.~\cite{Hayman25}).

The case of Fig.~\ref{fig:Tiles_A_B} discussed above is a particular case, where the boundaries between tiles are parallel to the $N_{\up}$ and $N_{\dw}$ axes, and therefore, while one KS potential jumps, the other one is continuous. In cases where the tile boundary is not parallel to either axis, both KS potentials will jump~\cite{BurgessLinscottORegan24},
and the magnitude of the corresponding $\Delta^\s$'s can be deduced in the same fashion as above.
In particular, when the boundary is not parallel to either axis, the jump in the KS potentials is not accompanied by $N_\s$ passing through an integer. This is in contrast to the usual derivative discontinuity of Ref.~\cite{PPLB82}.

Furthermore, at points on the boundary between two tiles, the values of $\eps_\ho^\s$ will match those of one of the two tiles. Which tile these belong to is determined by the fact that in Eq.~\eqref{eq:eps_ho_dE_dN} we must take the derivative with respect to $N_\s$ from below:
$\eps_\ho^\s= \left( \frac{\de E}{\de N_{\s}^-}\right)_{{N}_{\bar{\s}}}$.
For instance, in Fig.~\ref{fig:Tiles_A_B}, $\eps_\ho^\dw$ evaluated on the boundary between tiles Tr and A, will equal the value of $(\eps_\ho^\dw)_A$, which is evaluated directly to the left of the boundary, in Tile~A.

\section{Numerical Analysis} \label{sec:NumInvestigation}

For further analysis, we make use of experimental and computational data from the NIST Atomic Spectra Database~\cite{NIST_ASD}, for atoms with atomic number $1 \leqs Z \leqs 54$. We have written a numerical code, which extracts from the NIST databases the system's ground-state energy and separately the spin-flip energies, for each $Z$ and each $N \leqs Z$. The latter are taken from the list of excited levels, being the first excitations with a given spin, which is higher than that of the ground state.
This procedure results in the values of $E_\NM$, within the set of $N$ and $M$ available in the database, for each $Z$.  Next, for any given $N_\tot$ and $M_\tot$, the ensemble coefficients $\lambda_\NM$ can be computed, by checking each possible triad of points $(N,M)$, and picking the triad that minimizes the energy $\sum_{N}\sum_{M}\lambda_\NM E_\NM$ of the resulting ensemble, subject to the  constraints~\eqref{eq:lambda_0_1},~\eqref{eq:Tr_Lambda},~\eqref{eq:N_tot} and~\eqref{eq:M_tot}.

\begin{figure}
    \centering
    \includegraphics[width=\linewidth]{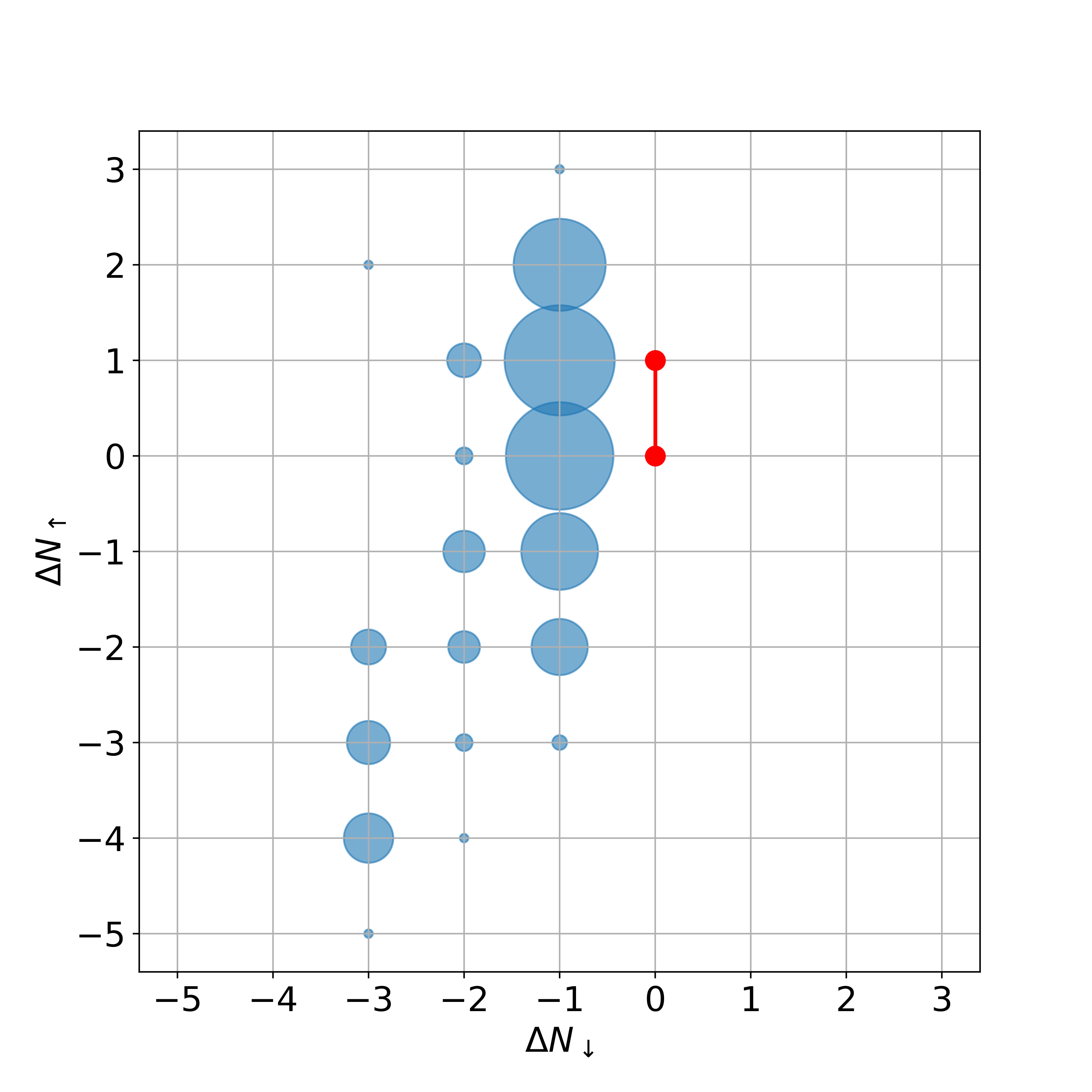}        \includegraphics[width=\linewidth]{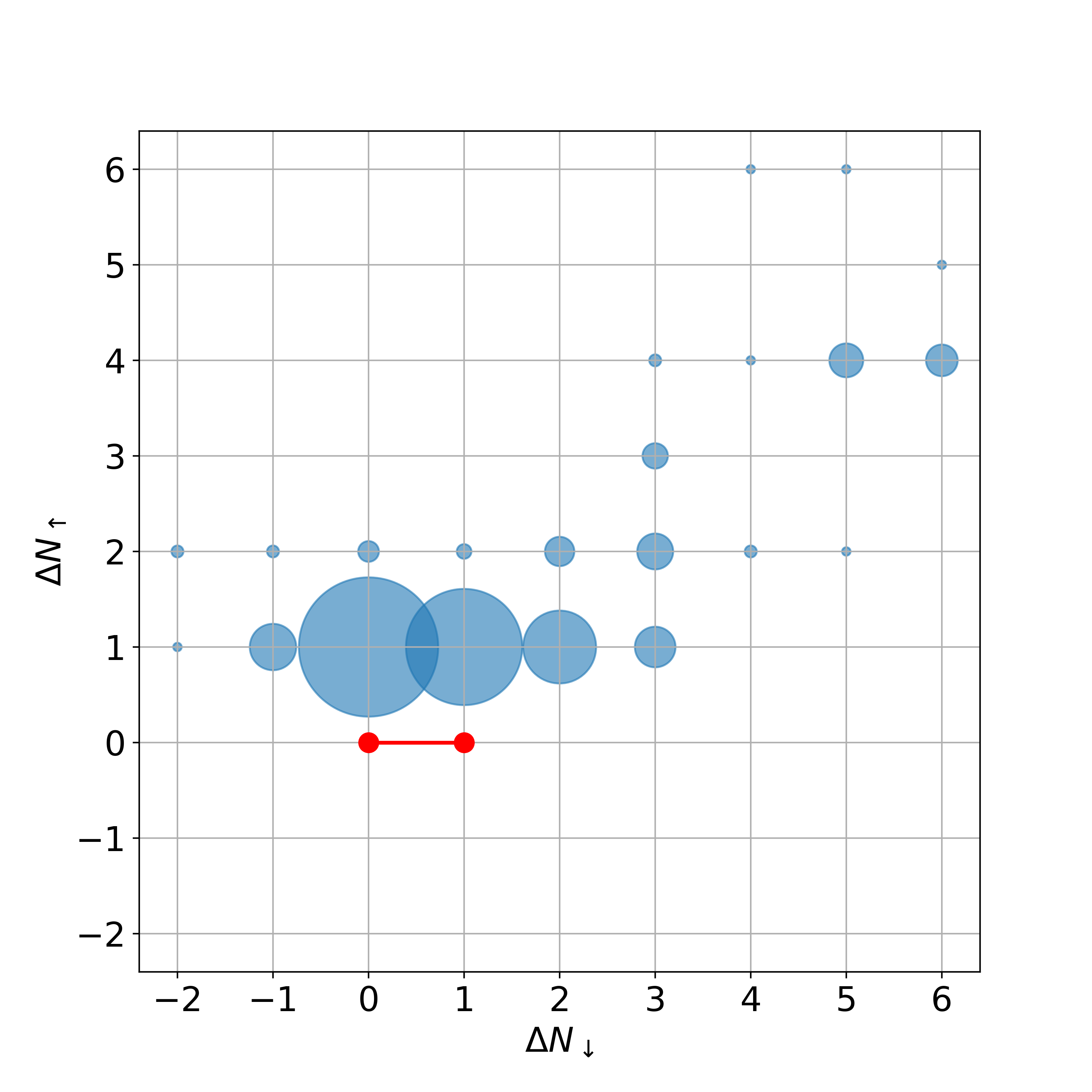}
    \vspace{0.5em}
    \caption{Abundance of the third state $(N_\star,M_\star)$ in the ensemble near the $M_B$ boundary, with respect to $(N_0,S_0)$, among atoms and ions with $1\leqs Z\leqs 54$, in the cases $S_1=S_0+\half$ (top) and $S_1=S_0-\half$ (bottom). The axes are $\Delta N_\s=N_{\star,\s}-N_{0,\s}$, where $N_{\star,\s}$ are the electron numbers at $(N_\star,M_\star)$, and $N_{0,\s}$ are the electron numbers at $(N_0,S_0)$. The area of each circle is proportional to the abundance of that $(\Delta N_\up, \Delta N_\dw)$. The red circles in each plot correspond to the two points on the boundary, $(N_0,S_0)$ and $(N_0 +1,S_1)$.}
    \label{fig:psi-star-stats}
\end{figure}

Furthermore, our code finds which points $(N_\star,M_\star)$
contribute to the ensemble at $M_{\tot}=M_B+
\delta$, when $\delta \rarr 0^+$, as we have discussed in Sec.~\ref{sec:Beyond_Reg_1}. This is achieved for each $N_0$ by minimizing the energy of Eq.~\eqref{eq:DE_star} over all possible points $(N_\star,M_\star)$. These results are presented and discussed below, in Fig.~\ref{fig:psi-star-stats}.

We wish to briefly comment on the limitations of our analysis. First, within the NIST database, usually only the first and sometimes the second spin-flip energies are available. In certain systems, no spin-flip energies appear at all.
This means that some $N,M$ values which may be relevant in our analysis, are missing. Furthermore, we include only levels which have been described within the $LS$-coupling scheme. Levels which are reported to include an unknown energy shift with respect to the rest of the spectrum are excluded.

\begin{figure}[htbp]
    \centering
    \includegraphics[width=0.85\linewidth]{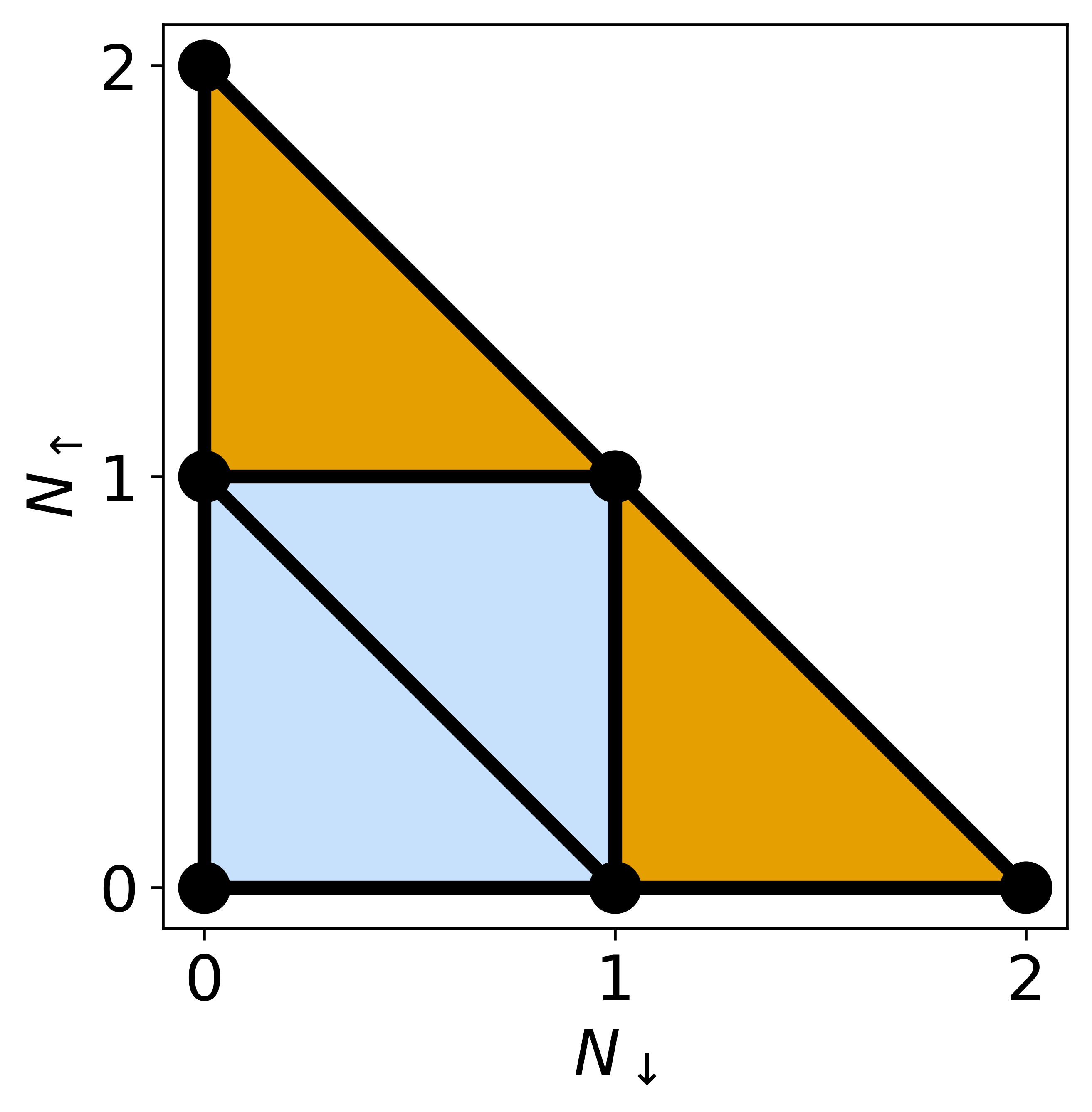}
    \caption{The ensemble plot for the He atom, where different colors indicate the set of pure states contributing to the ensemble at each $(N_\up,N_\dw)$. Region~\eqref{eq:spin condition} is shown in light blue. Lattice points whose corresponding pure state energy is available in the NIST database, are marked by a dot. }
    \label{fig:ens-plot-He}
\end{figure}

\begin{figure}[htbp]
    \centering
    \includegraphics[width=0.85\linewidth]{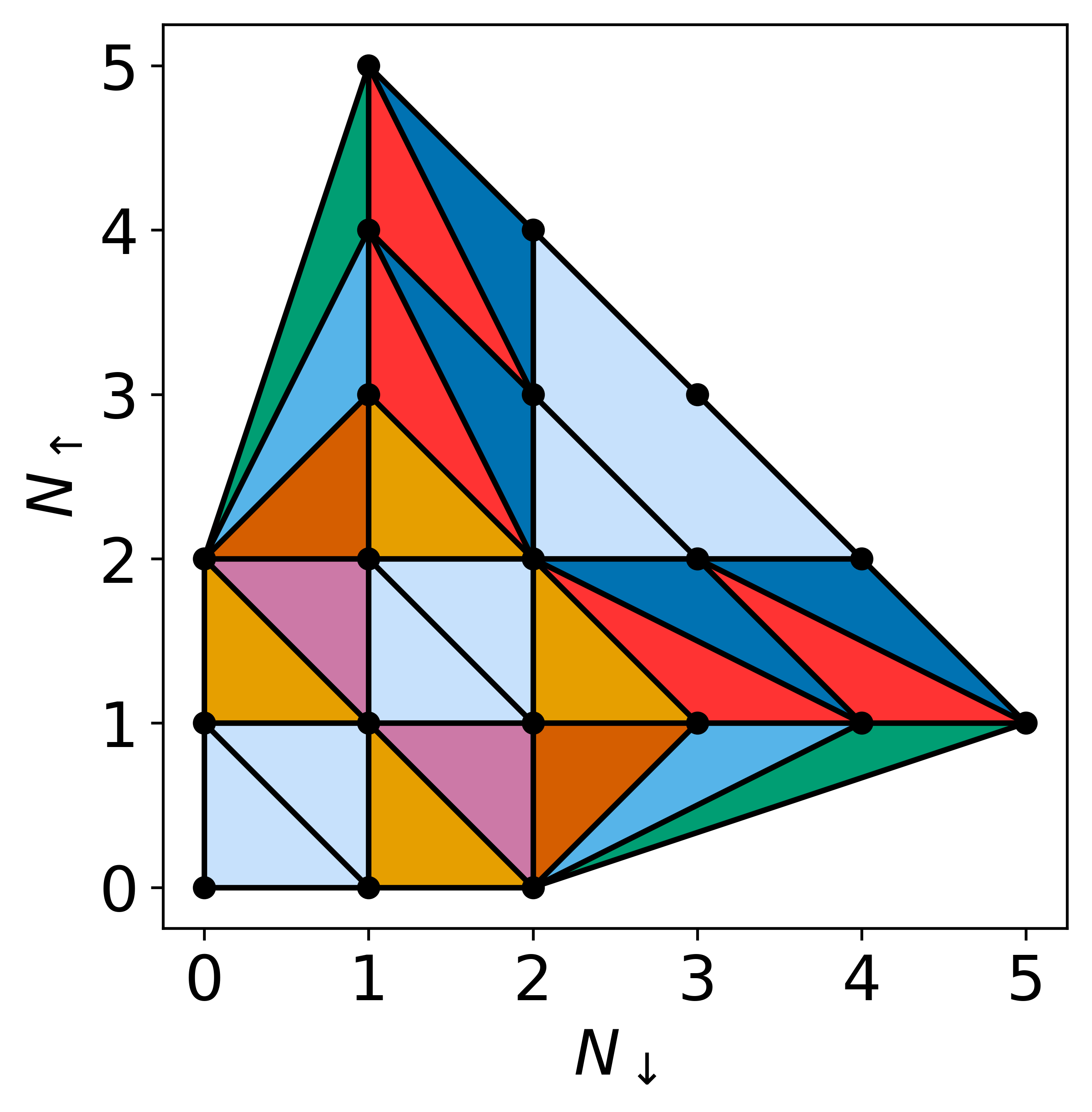}
    \caption{Same as Fig.~\ref{fig:ens-plot-He}, but for the C atom.}
    \label{fig:ens-plot-C}
\end{figure}

\begin{figure}
    \centering
    \includegraphics[width=0.98\linewidth]{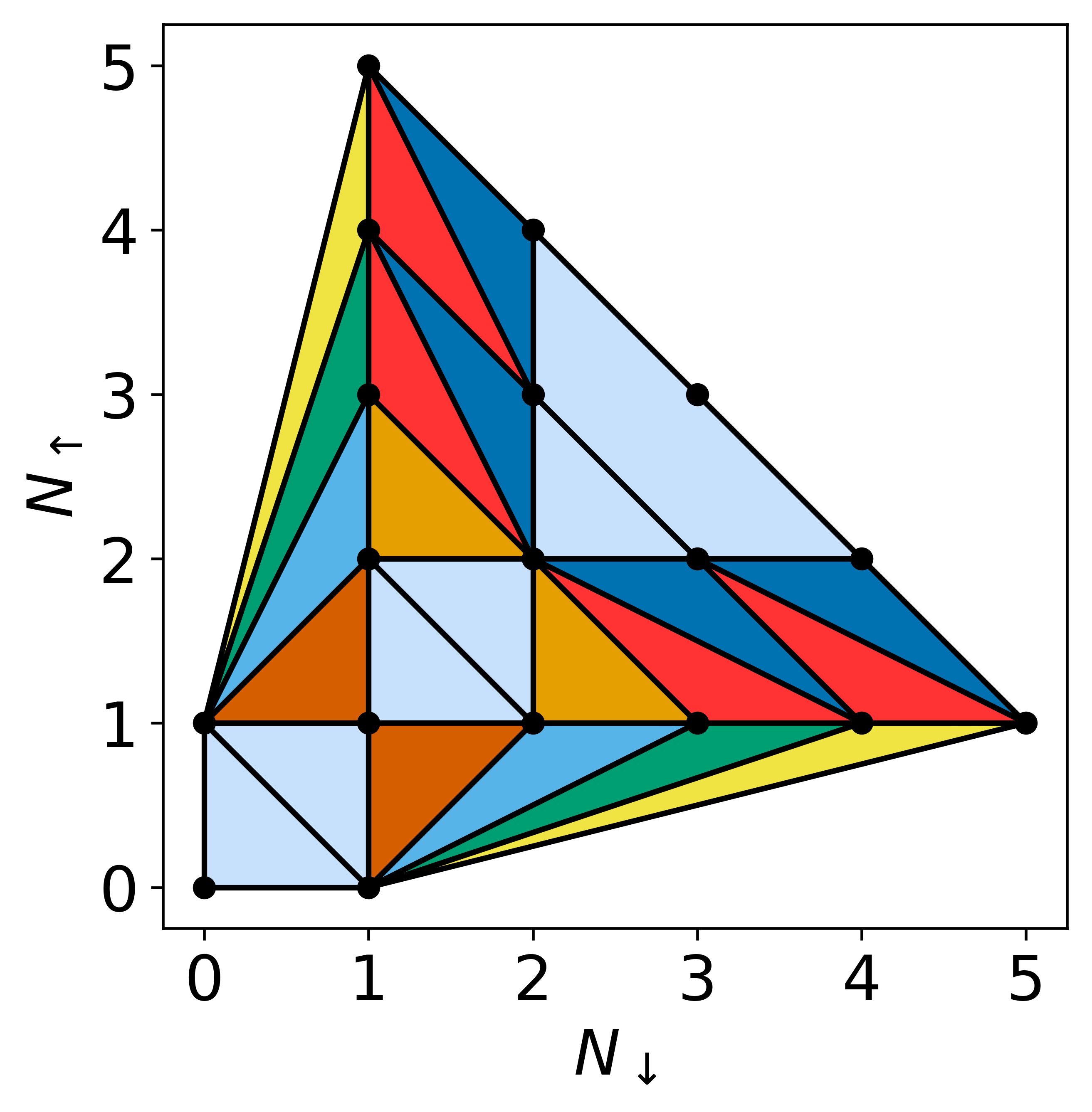}
    \caption{Same as Fig.~\ref{fig:ens-plot-C} (for the C atom), but excluding the point ($N_\up$,$N_\dw$)=(2,0), demonstrating the ensemble plot's dependence on the available pure state energies. Triangles which had a vertex at $(2,0)$ in Fig.~\ref{fig:ens-plot-C}, now have a vertex at $(1,0)$ instead.
    }
    \label{fig:ens-plot-C-exp-only}
\end{figure}

\begin{figure}[htbp]
    \centering
    \includegraphics[width=0.98\linewidth]{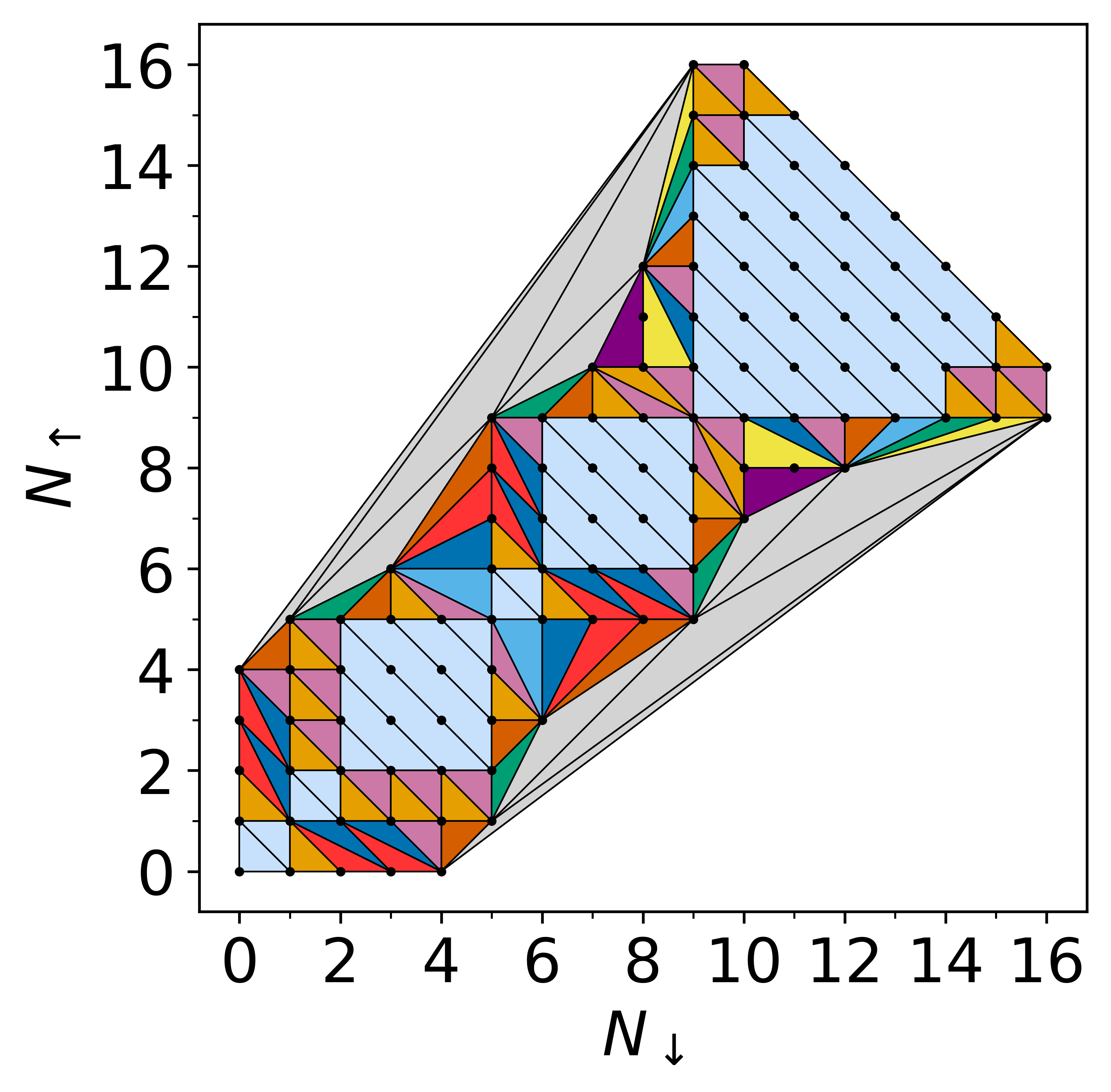}
    \caption{Same as Fig.~\ref{fig:ens-plot-C}, but for the Fe atom. Due to a lack of data for highly excited states, some tiles (shown in gray) appear stretched, being obvious artifacts of missing $(N,M)$ points.}
    \label{fig:ens-plot-Fe}
\end{figure}

Using the values of $E_\NM$, our code determines the pure states constituting the ensemble at each point in the $N_\up - N_\dw$ plane. In this way, we obtain the \emph{ensemble plots} for various systems, presented in Figs.~\ref{fig:ens-plot-He}--\ref{fig:ens-plot-Fe}.
Being an extension of Fig.~\ref{fig:trapezoids}, these present the tiles in the $N_\up - N_\dw$ plane in color, indicating which pure states contribute to a given point $(N_\up,N_\dw)$. For all points belonging to the same triangular tile, the contributing pure states are the vertices of that tile. When exactly on a tile boundary, the contributing states are those belonging to this boundary.

The simple example of the He atom is depicted in Fig.~\ref{fig:ens-plot-He}. Here, all the points with $0<N<1$ are described by an ensemble that comprises of $(N_\up,N_\dw)=(0,0)$, $(1,0)$ and  $(0,1)$. Points with $1<N<2$ and $-M_B<M<M_B$ are described by an ensemble that comprises of $(N_\up,N_\dw)=(1,0)$, $(0,1)$ and $(1,1)$. These are within Region~\eqref{eq:spin condition} (light blue).  Outside Region~\eqref{eq:spin condition} (orange triangles), points with $1<N<2$ and $M > M_B$ are described by an ensemble of the pure states at $(N_\up,N_\dw)=(1,0), (1,1)$ and $(2,0)$. For $M < -M_B$, the relevant pure states are $(N_\up,N_\dw)=(0,1), (1,1)$ and $(0,2)$.

Next, Fig.~\ref{fig:ens-plot-C} shows the example of the C atom. In addition to the already familiar triangles and trapezoids (light blue; within Region~\eqref{eq:spin condition}), we observe triangular tiles of various types. Here we wish to emphasize the following points:

(i)~Alongside the `conventional' right triangle tiles, (orange, e.g. the vertices (2,1), (2,2), (3,1) and brown, e.g., the vertices (2,1), (3,1), (2,0)), we observe the blue and red tiles (e.g., (3,2), (4,2), (5,1) and (3,2), (4,1), (5,1), respectively), which exemplify the fact that the ensemble is not always comprised of states that belong to the smallest square enclosing the ensemble point $(N_\up,N_\dw)$ (cf.~\cite{XDYang16}).
The fact that the blue tile is not conventional, i.e.\ we did not get the right triangle (3,2), (4,2), (4,1), is directly related to he fact that the first spin-flip energy of neutral C, $J_{1,6} = 4.18$~eV is \emph{lower} than the first spin-flip energy of C$^+$, $J_{1,5} = 5.33$~eV~\cite{NIST_ASD} (cf.\  Sec.~\ref{sec:CasesPsiStar})

(ii)~Crossing from the blue to the red tile results in a discontinuity in the energy profile, while $N_\up$, $N_\dw$, $N_\up+N_\dw$ and $N_\up-N_\dw$ are all fractional (shown also in Ref.~\cite{BurgessLinscottORegan24}); this situation is, in a sense, a result of (i).  This is in contrast to the well-established derivative discontinuities of Ref.~\cite{PPLB82} which occur at integer $N_\tot$, and also in contrast to the spin-migration derivative discontinuities at $M_\tot=\pm M_B$~\cite{MoriS09,Hayman25}.

(iii)~Even when varying $M_\tot$ for an \emph{integer} $N_\tot$ (spin migration scenario~\cite{Hayman25}), particularly for $N_\tot=4$, we observe a discontinuity in the energy slope at fractional $N_\up$ and $N_\dw$ (transition from the skyblue tile to the green tile at $(N_\up,N_\dw)=(3.5,0.5)$). Both in cases (ii) and (iii), we expect a discontinuous jump in $v_\KS^\up(\rr)$ and $v_\KS^\dw(\rr)$ simultaneously, as we cross the boundary. In fact, the scenario of a jump in \emph{only one} of the $\s$-KS potentials, which we frequently encountered in the past, occurs only when we cross a boundary that is parallel to one of the axes in the $N_\up-N_\dw$ plane.

(iv)~The absence of data for the second spin flip energies of C, C$^+$ and C$^{2+}$, and the spin flip energy of C$^{3+}$, leads to missing integer points at $(N_\up,N_\dw)=$ (3,0), (4,0) (5,0) and (6,0). If total energies for these points were available, the ensemble map could look different. In particular, the green tile may be replaced by triangles which include the point (4,0), (5,0) or (6,0).
This illustrates one of the limitations of our approach, which we mentioned above. However, addition of these missing points is unlikely to affect the red and blue tiles of C, and it certainly cannot replace them by conventional right triangles.

(v) To illustrate the dependence of the ensemble plots on the available $(N,M)$ points, in Fig.~\ref{fig:ens-plot-C-exp-only} we present an ensemble plot for the C atom where the point $(2,0)$ is excluded. The result is that each triangle which previously had $(2,0)$ as a vertex in Fig.~\ref{fig:ens-plot-C}, now has a vertex at $(1,0)$ instead. Therefore, augmentation of the existing $E_\NM$-database with further results is highly desirable.

\begin{figure}
    \centering
    \includegraphics[width=0.98\linewidth]{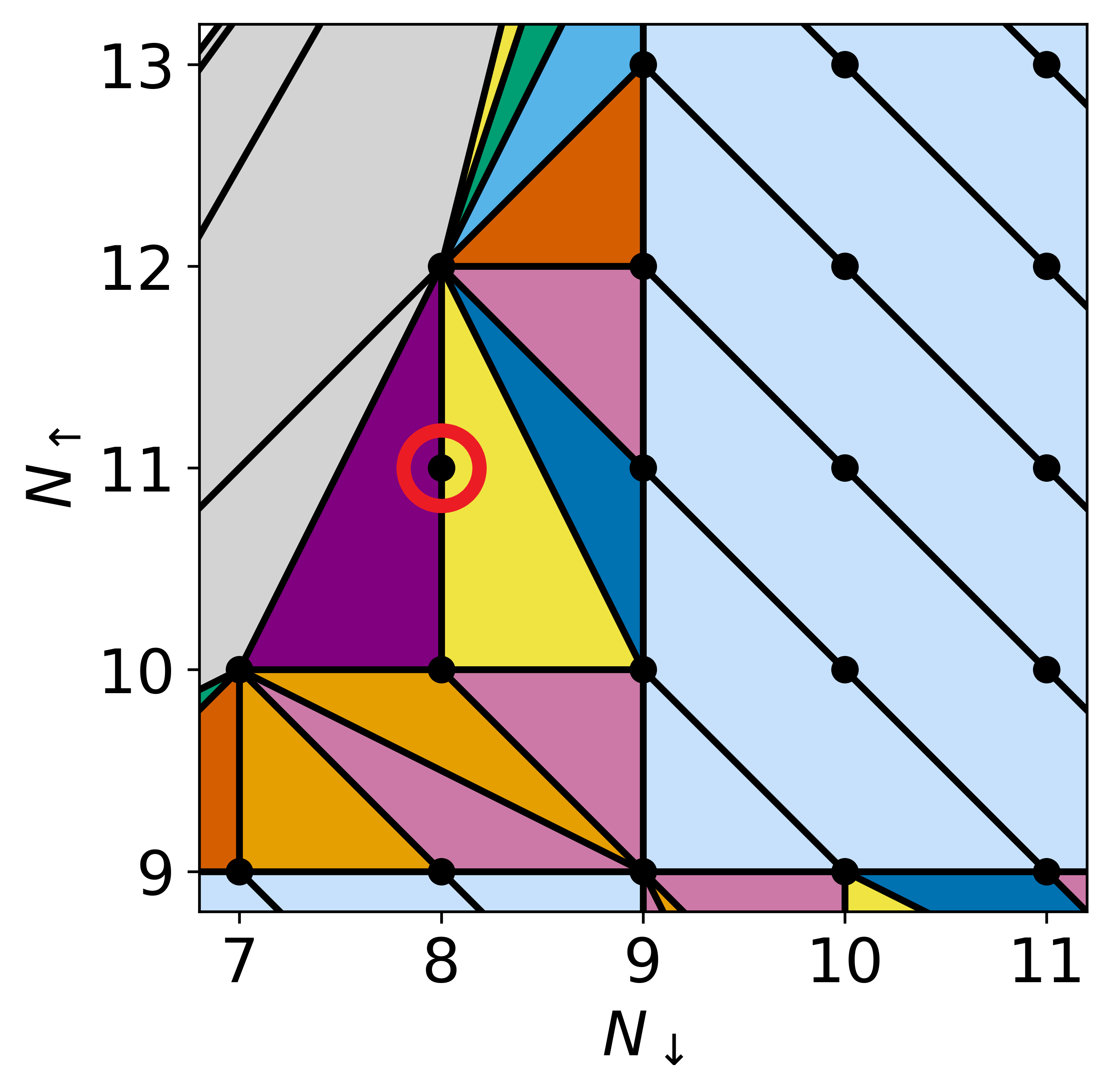}
    \caption{Same as Fig.~\ref{fig:ens-plot-Fe}, but zoomed-in near the point ($N_\up$,$N_\dw$)=(11,8), where the ensemble is not a pure state, but rather consists of (12,8) and (10,8).}
        \label{fig:nonconv-Fe}
\end{figure}

In Figs.~\ref{fig:ens-plot-Fe} and~\ref{fig:nonconv-Fe} we examine the case of Fe.
In addition to observing a multitude of non-conventional tiles as in C, we focus on the region around  $(N_\up,N_\dw)=(11,8)$ (Fig.~\ref{fig:nonconv-Fe}).
Surprisingly, at this point the ground state is an ensemble comprised of the states at (10,8) and (12,8), with coefficients of $\half$, and not the pure state (11,8)!  The aforementioned ensemble creates a lower energy than the pure state in question. If taken at face value, this suggests that the ion Fe$^{7+}$ with average spin 3/2 does not materialize as a ground state with 11 electrons up and 8 down. Instead, the system prefers a mixture of (10,8) and (12,8) ions. In this peculiar case, the set of pure state energies $E_\NM$ is not convex with respect to $N_\up$. Therefore, when ionizing an up-electron from the (12,8) ion, the spin-up ionization potential $I_\up$ decreases.
This result has been double-checked by us, and the conclusion is beyond the published experimental uncertainty: the ensemble energy is lower than the pure state energy by~43 eV, whereas the uncertainty in this difference is~$\pm 22$ eV. An extensive search for further examples of integer points outside the trapezoids, where the ground-state is not a pure state, did not yield additional results beyond the experimental/computational errors. However, there are several such examples, which are within the error range. The atoms and integer points where this occurs are Ga at (16,14), Se at (16,14), Kr at (16,14), Y at (19,17), In at (25, 23), and Sn at (25, 23).
Notably, all these examples are instances where the energy is not convex with respect to $N_\s$.  We did not find cases where the energy is not convex with respect to $M$ or, for this matter, other kinds of non-convexities, namely an instance of an integer point in the $N_\up - N_\dw$ plane, where a three-state ensemble yields an energy that is lower than the pure state of this point.

Finally, we numerically analyze the question of the ensemble near the border $M_\tot = M_B$ (see Sec.~\ref{sec:Beyond_Reg_1}), looking for the third state $(N_\star, M_\star)$ in the ensemble. We distinguish two cases: (i) $S_1 = S_0 + \half$ and (ii) $S_1 = S_0 - \half$.  Fig.~\ref{fig:psi-star-stats} presents the abundance of different $(N_\star, M_\star)$, relative to the point $(N_0, S_0)$. The area of each circle in these figures is proportional to the abundance of the corresponding state in the dataset we examined. We find that the most common cases are $(N_\star, M_\star) = (N_0, S_0+1)$ and $(N_0-1, S_0+\half)$ for~(i) and  $(N_\star, M_\star) = (N_0+1, S_0+\half)$ and $(N_0+2, S_0)$ for~(ii). These create the conventional, right triangles in the $N_\up-N_\dw$ plane. Ensembles with points farther from the boundary are progressively less common. Note that states that lie far from the boundary $M_B$ could be artifacts related to the lack of experimental and/or computational data we mentioned above.

\section{Discussion and conclusions} \label{sec:Conculsions}

In this article, we analyzed the fundamental question of what is the ensemble ground state of a general, finite, many-electron system, with given, possible fractional values of the electron number and the spin. We addressed both the case of low spin, namely when $M_\tot$ is within Region~\eqref{eq:spin condition}, and the complementary case of high spin.

When the spin is within Region~\eqref{eq:spin condition}, the exact ensemble form for the ground state has been found in Ref.~\cite{GoshenKraisler24}. In Sec.~\ref{sec:in_Reg_1} we provide an alternative derivation for this result.
Furthermore, we discuss the question of the ambiguity in the ensemble ground state, exemplify it by analyzing the spin polarization $\zeta(\rr)$ and show that the requirement of maximum entropy~\eqref{eq:S_def} defines the ensemble ground state uniquely. Notably, maximizing the entropy leads to a ground state that is generally different from the one obtained by minimizing $\Delta S_z$ or by applying a small magnetic field.

It is worth mentioning that since we are dealing with systems at zero temperature, namely with solving the quantum-mechanical problem for the many-electron Hamiltonian, without invoking thermal ensembles, the requirement of maximal entropy is not compulsory. Strictly speaking, it is external to the theoretical framework in which the problem has been formulated. Nonetheless, entropy maximization is useful to remove the ground state ambiguity, as we showed; other ways exist, as well. Still, when looking at many-electron systems at finite temperature, our task would be to minimize the free energy $F = E - T\mathscr{S}$, which in the limit $\kB T \rarr 0^+$, would mean to minimize $E_\tot$ and then maximize $\mathscr{S}$, as performed in Sec.~\ref{sec:in_Reg_1}.
\blue{In fact, for a collection of many replicas of a system, with a specified average electron number and average spin, allowing them to exchange $\up$ and $\dw$ electrons, the ensemble state of the system will be the statistically likeliest ground state, i.e., the one with maximal $\mathscr{S}$. In this context, the physical choice is the maximum entropy ensemble.}


In contrast to the situation within Region~\eqref{eq:spin condition}, for the high-spin case, the form of the ensemble ground state strongly depends on the system in question. In Sec.~\ref{sec:Beyond_Reg_1} we succeeded to prove three general properties, which characterize the unknown ground state and narrow down the list of pure states it may consist of. These properties are: (1) For $M_\tot > M_B$, the ensemble consists only of pure states which have $M \geqs S_{\m}(N)$.
(2) The ensemble consists of at most three pure states.
(3) Near the boundary $M_\tot=M_B$, with $N_\tot$ strictly fractional, two of the pure states in the ensemble are known, and the third depends on the particular system, but is independent of $\alpha$.
To illustrate the properties of high-spin ensembles close to the boundary $M_\tot=M_B$, in Sec.~\ref{sec:CasesPsiStar} we discussed in detail two particular cases: $S_1 = S_0 + \half$ (addition of an up electron to an $N_0$-electron system) and $S_1 = S_0 - \half$ (addition of a down electron). We considered common scenarios for $\ket{\Psi_\star}$, compared their energies and connected the latter to the IPs, fundamental gaps and spin-flip energies of these systems. Lastly, using the examples we developed above, we illustrated how the case of accidental degeneracy mentioned in Sec.~\ref{sec:Beyond_Reg_1} could emerge. All the aforementioned results are general properties of many-electron systems, and are not related to a specific method chosen for their description.

Next, in Sec.~\ref{sec:eigenvalues} we focused on exact KS-DFT and related the KS frontier eigenvalues $\eps_\ho^\s$ to total energy differences, particularly the IP and the spin flip energies. The Li atom served as a useful illustration.  In this way, in Eqs.~\eqref{eq:e_ho_s_I}, \eqref{eq:e_ho_s_IJ_Tile_A}--\eqref{eq:e_ho_s_IJ_Tile_B_prime}, we derived an extension of the well-known IP theorem to systems with any $N_\tot$ and $M_\tot$.
Furthermore, when crossing the boundary between any two tiles in the total energy profile, either one or both of $\eps_\ho^\s$ jump discontinuously. This is accompanied by a derivative discontinuity $\Delta^\s$, which manifests as a jump in the corresponding potential $v_\KS^\s(\rr)$. 
These new exact conditions for the orbital energies $\eps_\ho^\s$, discontinuities $\Delta^\s$ and potentials $v_\KS^\s(\rr)$ should be obeyed by approximate xc functionals in DFT and will be useful for the development of advanced approximations. 




In Sec.~\ref{sec:NumInvestigation}, we performed an extensive analysis of existing experimental and computational data~\cite{NIST_ASD}, from which we obtained the pure state energies $E_\NM$ in atomic systems. Relying on this analysis, we created the ensemble maps for various atomic systems, which show the states that contribute to the ensemble ground state, for given $N_\tot$ and $M_\tot$, in practice. We found that the ensemble maps consist of trapezoids and triangles of different types. Particularly, the ensemble is not always comprised of states that belong to the smallest square enclosing the ensemble point $(N_\up,N_\dw)$, both for low and high spin. Furthermore, discontinuities may appear at any integer or fractional $N_\up$ and/or $N_\dw$, in one of the spin channels or in both; reproducing these exact properties is a significant challenge for xc approximations of the next generation.
Next, we performed a statistical analysis as to the occurrence of the third state $(N_\star, M_\star)$, when close to the boundary spin $M_B$. Lastly, we examined the convexity of the discrete pure state energy function, $E_\NM$. We found that $E_\NM$ is nearly always a convex function (with the exception of Fe$^{7+}$). Particularly, this supports the assumption of convexity with respect to $N$, to $M$ and to $N_\s$. Notably,
in Ref.~\cite{Hayman25},
$M$-convexity proved useful for investigation of spin migration in atomic systems. 

The new exact, general properties of many-electron systems with a varying, possibly fractional, electron number and varying spin, both in the low- and high-spin regions, derived in this work, are \blue{expected to be useful in the broad context of materials research, in several respects. At the first stage, exact properties serve as points of reference to which approximate computational results can be compared, and the quality of approximation can be assessed. Such a comparison is relevant for DFT results~\cite{Prokopiu22,Hayman25}, as well as for results emerging from wavefunction-based or Green-functions-based methods.  At the second stage, exact results can serve to correct approximate DFT functionals by tuning free parameters, e.g., within the line of thought of optimally tuned range-separated hybrids~\cite{BNL1,BNL2,Baer10,Kronik_JCTC_review12}. At the third stage, our findings can serve to develop advanced approximations: either by derivation of an entirely new exchange-correlation functional with the new exact properties serving as exact constraints~\cite{Perdew09}, by proposing an ensemble-based generalization (in the spirit of,  e.g., Refs.~\cite{KraislerKronik13,Cernatic24,Gould_Ensemblization26}) or by incorporating exact constraints via physically-informed machine-learning techniques~\cite{DM21,Callow23,Cangi24,MALA25}. In all cases, improvement in the predictive power of electronic structure methods from first principles, especially for spin-related properties, is to be expected.}


\section*{Supplementary Material}
Technical details which support our findings are available in the SM. These include numerical details on calculations of the spin-polarization $\zeta(\rr)$ for the C atom, and further mathematical arguments which are not presented in the main text.

\acknowledgments

We thank Yevgeny Rakita, Sangeeta Sharma and Sam Shallcross for fruitful discussions.

\bibliography{bib2023}
\end{document}


\title{Supplementary Material for: \\ \textit{Many-electron systems with fractional electron number and spin: \\ exact properties above and below the equilibrium total spin value}}

\author{Yuli Goshen}
\affiliation{Fritz Haber Research Center for Molecular Dynamics and Institute of Chemistry, The Hebrew University of Jerusalem, 9091401 Jerusalem, Israel}

\author{Eli Kraisler}
\email[Author to whom correspondence should be addressed: ]{eli.kraisler@mail.huji.ac.il}
\affiliation{Fritz Haber Research Center for Molecular Dynamics and Institute of Chemistry, The Hebrew University of Jerusalem, 9091401 Jerusalem, Israel}

\date{\today}

\begin{abstract}
In this document we provide additional, largely technical details, which support the findings presented in the main text of the article. 
\end{abstract}

\maketitle

The Supplemental Material presented below provides additional details on various topics discussed in the main text (MT). When referring to an equation or a figure from the main text, its number is preceded by the letters MT. Equation and figure numbers without letters refer to this document.

In Sec.~\ref{sec:zeta}, we provide technical details on calculations of the spin polarization $\zeta(\rr)$ for the C atom with fractional electron number, and present additional numerical results for this quantity.
In Sec.~\ref{sec:minDSz}, we show that  within Region~\EqSpinCond, minimization of $\Delta S_z$ leads to a unique ensemble, whose constituents correspond to three of the four vertices of the square that encloses the point $(N_\tot,M_\tot)$ in the $N_\up-N_\dw$ plane.
We have presented a proof for the uniqueness of the minimum $\Delta S_z$ ground state already in Ref.~\cite{GoshenKraisler24}, and here a different, improved argument is given, indicating which states comprise the ensemble.
In Sec.~\ref{sec:maxS}, we show that within Region~\EqSpinCond, the ground state which maximizes the entropy is one which includes every allowed pure state with a nonzero weight, as required in our derivation of the unique maximum-entropy ground state.
In Sec.~\ref{sec:ContDiff_Eq39}, we show that the left-hand side (lhs) of Eq.~\SmaxMtotCondition~is a continuous, differentiable, and strictly increasing function, implying the uniqueness of the solution to Eq.~\SmaxMtotCondition, as required.
Finally, in Sec.~\ref{sec:InternalDeg} we provide details about the maximal entropy ensemble in the case of internal degeneracy.

\section{Spin polarization of the C atom} \label{sec:zeta}
For the C atom with $5\leqs N_\tot\leqs6$ electrons, within Region~\EqSpinCond, the spin polarization $\zeta(\rr)$ is given by Eq.~\EqZetaCarbon, where $S_0=\thalf$ and $S_1=1$. The polarization has one free parameter, $y$, due to the ambiguity of the ensemble ground state, as explained in the main text. Although undetermined, the parameter $y$ is constrained: $|y| \leqs \thalf (1-\alpha)$ and $|y-M_\tot| \leqs \alpha - x$. These constraints stem from the fact that all the statistical weights in the ensemble must be within $[0,1]$.

To graphically illustrate the influence of the ground-state ambiguity, we performed KS-DFT  calculations for neutral C and for its first ion C$^+$. All calculations were carried out with the \texttt{ORCHID} atomic
spherical DFT code~\cite{KraislerMakovArgamanKelson09,KraislerMakovKelson10}, using a logarithmic grid of $N = 16,000$ points
on the interval $[r_\textrm{min}, r_\textrm{max}]$, where $r_\textrm{min} = e^{−13}/Z$ Bohr, with $Z=6$ being the
atomic number, and $r_\textrm{max} = 35$~Bohr. We performed calculations with four exchange-correlation (xc) functionals: the local spin-density approximation (LSDA)~\cite{PW92}, the Perdew-Burke-Ernzerhof (PBE) generalized gradient approximation~\cite{PBE96}, the exact exchange (EXX) and the hybrid PBE0 functional~\cite{AdamoBarone99}. Calculations for EXX and PBE0 were perfomed within the Krieger–Li–Iafrate (KLI)~\cite{KLI92} approximation to the Optimized Effective Potential (OEP) method~\cite{KK08}, remaining strictly within the KS formalism.

In all cases, for C$^+$ with $M=S_0=\thalf$, we obtained the densities $n^\up_{N_0,\half}(r)$ and $n^\dw_{N_0,\half}(r)$, and for neutral C with $M=S_1=1$, we obtained $n^\up_{N_0+1,1}(r)$ and $n^\dw_{N_0+1,1}(r)$. These are sufficient to construct $\zeta(r)$, according to Eq.~\EqZetaCarbon ~[To emphasize, the densities $n^\s_{N_0+1,0}(r)$ were set to be  $n^\up_{N_0+1,0}(r) = n^\dw_{N_0+1,0}(r) = \thalf(n^\up_{N_0+1,1}(r)+n^\dw_{N_0+1,1}(r))$, as for the exact case, and were not obtained separately, e.g., from a calculation with $M=0$].

\begin{figure}
    \centering
\includegraphics[width=0.98\linewidth]{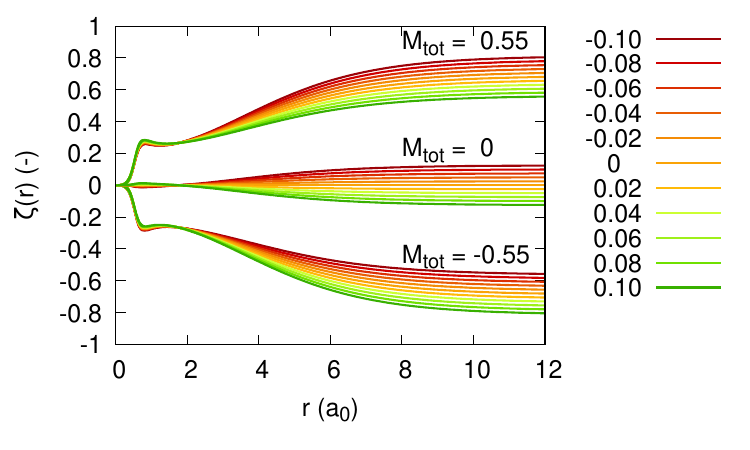}
\caption{Spin polarization $\zeta(r)$, for the C atom, with $\alpha=0.8$ and $M_\tot=0.55$, $0$ and $-0.55$ (see graph), for various values of the parameter $y$ defined in the main text (see Legend), calculated with the LSDA}
    \label{fig:C__zeta_M}
\end{figure}

Figure~\FigZetaCarbon~shows $\zeta(r)$ for C with 5.8 electrons ($\alpha=0.8$), with $M_\tot=0$, for various values of $\alpha$, calculated within the LSDA. We clearly see that $\zeta(r)$ depends on $y$ dramatically; particularly, far from the origin, $\zeta(r)$ reaches the limit $-y/(\alpha S_1)$. This is due to the fact that in our example $n^\up_{N_0+1,1}(r)$ is the density with the slowest decay rate.
Furthermore, Fig.~\ref{fig:C__zeta_M} here presents $\zeta(r)$ also for $M_\tot=0.55$ and $-0.55$. The dependence on the parameter $y$ remains significant, with the asymptotic limit of $(M_\tot-y)/(\alpha S_1)$.

\begin{figure}
    \centering
\includegraphics[width=0.98\linewidth]{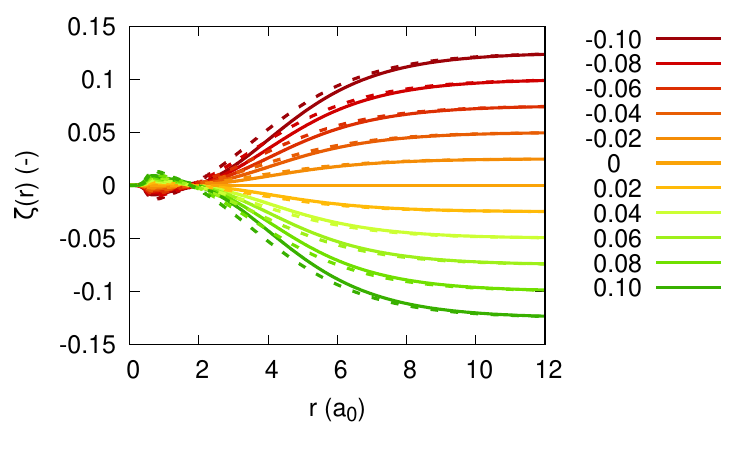}
\includegraphics[width=0.98\linewidth]{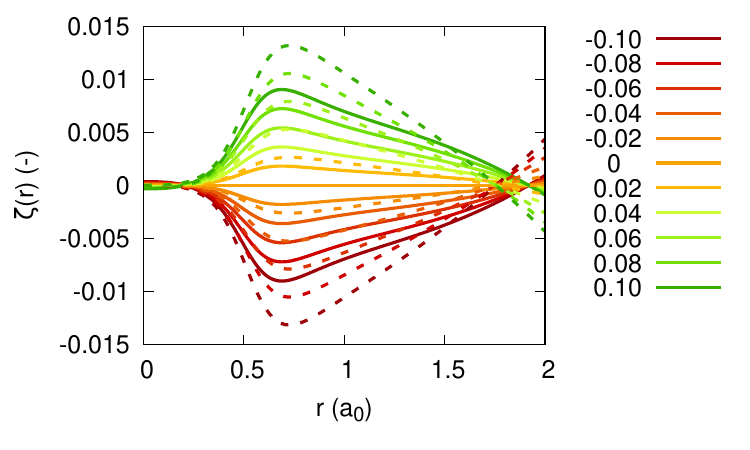}
\caption{Top: Spin polarization $\zeta(r)$, for the C atom, with $\alpha=0.8$ and $M_\tot=0$, for various values of the parameter $y$ defined in the main text (see Legend), comparing EXX results (solid lines) and LSDA results (dashed lines). Bottom: same as top panel, focusing on $r <2$ Bohr.}
    \label{fig:C__zeta_M_0_EXX_LSDA}
\end{figure}

Calculations with other xc functionals provide polarization functions that may somewhat differ in their details, but are qualitatively the same. Figure~\ref{fig:C__zeta_M_0_EXX_LSDA} compares results for EXX-KLI vs.\ LSDA. We find that whereas the exact polarization curves do not strictly overlap, and closer to the origin (bottom panel) the differences between LSDA and EXX curves are clearly visible, the result is overall very simiar; particularly, the curves for both approximations have exactly the same asymptotics, as expected.

Comparison of four sets of approximate DFT results suggests that also if we were using exact electron densities, we would be arriving at the same conclusion we stated in the main text: the spin polarization is not uniquely set, does \emph{not} depend on $x$ and does strongly depend on $y$. The asymptotic limit for the spin polarization is $\zeta_{\textrm{asymp}} = (M_\tot-y)/(\alpha S_1)$, as long as the slowest decay is for $n^\up_{N_0+1,1}(\rr)$.

\section{the Minimum $\Delta S_z$ ground state} \label{sec:minDSz}

In this Section, we aim to derive the ground state within Region~\EqSpinCond, which minimizes $\Delta S_z=\sqrt{\stat{\hat{S}^2_z}-\stat{\hat{S}_z}^2}$,
under the constraints
\EqconstraintRange, ~\EqconstraintOne, ~\EqconstraintTwo ~and \EqconstraintThree. We first notice the identity
\begin{equation}
    (\Delta N)^2 + (2 \Delta S_z)^2=2 ((\Delta N_\up)^2+(\Delta N_\dw)^2),
    \end{equation}
where
$(\Delta N_\s)^2 =
\stat{\hat{N}_\s^2}
-\langle \hat{N}_\s \rangle^2$ is the variance in the number of spin-$\s$ electrons,
and likewise
$(\Delta N)^2 =
\stat{\hat{N}^2}-
\langle \hat{N}\rangle ^2$. Within Region~\EqSpinCond, namely when the ground state is given by Eq.~\EqGammaTrapezoids, it can be shown that $(\Delta N)^2 = \alpha(1-\alpha)$. Since the latter quantity is completely determined by the value of $N_\tot$ and is independent of the specific ensemble weights, minimizing $\Delta S_z$ is equivalent to minimizing $(\Delta N_\up)^2+(\Delta N_\dw)^2$.
We will show that $(\Delta N_\up)^2$ and $(\Delta N_\dw)^2$ can be minimized simultaneously, leading to the global minimum of $\Delta S_z$.

We now express the ensemble {of Eq.~\EqGammaTrapezoids} as
\begin{equation}\label{eq:Lambda_updw}
\hat{\Lambda}=\sum_{n_{\uparrow}n_{\downarrow}}\lambda_{n_{\uparrow}n_{\downarrow}}\ketbra{\Psi_{n_{\uparrow}n_{\downarrow}}}{\Psi_{n_{\uparrow}n_{\downarrow}}},
\end{equation}
where the new indices $n_\s$ indicate
the number of \mbox{spin-$\s$} electrons in the pure state $\ket{\Psi_{n_\up n_\dw}}$.
Equation~\eqref{eq:Lambda_updw} is the same as Eq.~\EqGammaTrapezoids, with the same pure states and ensemble weights, only we now index them using $n_\up$ and $n_\dw$, rather than $N$ and $M$.

In the following, we describe the minimization of $(\Delta N_\up)^2$ and $(\Delta N_\dw)^2$. The argument is identical for both, hence  we now focus on just the $\up$ channel. We express $\left(\Delta N_{\uparrow}\right)^{2}=\sum_{n_{\uparrow}}\lambda_{n_{\uparrow}}n_{\uparrow}^{2}-\left(\sum_{n_{\uparrow}}\lambda_{n_{\uparrow}}n_{\uparrow}\right)^{2}$, where we have denoted $\lambda_{n_{\uparrow}}=\sum_{n_{\downarrow}}\lambda_{n_{\uparrow}n_{\downarrow}}$.
We shall now find the set of $\left\{ \lambda_{n_{\uparrow}}\right\} $, which minimize $\left(\Delta N_{\uparrow}\right)^{2}$, while satisfying  $N_{\uparrow}=\sum_{n_{\uparrow}}\lambda_{n_{\uparrow}}n_{\uparrow}$, as well as $\sum_{n_\up} \lambda_{n_\up} =1$ and $\lambda_{n_{\uparrow}}\in\left[0,1\right]$. In this constrained minimization, only the coefficients $\lambda_{n_\up}$ appear rather than $\lambda_{n_{\up} n_\dw}$.

In the spirit of Sec.~II\thinspace A of the main text, applying constrained minimization with Lagrange multipliers $w_1$ and $w_2$, we have $n_{\uparrow}^{2}+\left(w_{1}-2N_{\uparrow}\right)n_{\uparrow}+w_{2}=0$ for each $n_\up$, such that $\lambda_{n_{\uparrow}}\neq0,1$.
This quadratic equation in $n_\up$ has at most two integer solutions, $u$ and $v$ (if there is a single solution, then $u=v$).
Without loss of generality, let us set $v \leqs u$.

Left with only two coefficients $\lambda_u$ and $\lambda_v$, our two constraints are simplified to
\begin{equation}\label{eq:uvConstraint}
\lambda_{u}u+\lambda_{v}v=N_\up
\end{equation}
and
\begin{equation} \label{eq:uvProbability}
    \lambda_u+\lambda_v=1.
\end{equation}
Using Eqs.~\eqref{eq:uvConstraint} and \eqref{eq:uvProbability}, we eliminate $\lambda_u$ and $\lambda_v$ in the expression for $(\Delta N_\up)^{2}$, to get

\begin{equation}\label{eq:uvVariance}
    \left(\Delta N_{\uparrow}\right)^{2}= (\lambda_u u^2+ \lambda_v v^2)-(\lambda_u u + \lambda_v v)^2=\left(N_{\uparrow}-v\right)\left(u-N_{\uparrow}\right).
\end{equation}
Since the lhs of Eq.~\eqref{eq:uvVariance} is positive, it is clear that $v \leqs N_\up \leqs u$. We then realize that $\left(\Delta N_{\uparrow}\right)^{2}$ as expressed in Eq.~\eqref{eq:uvVariance} is minimized only by $v=\text{floor}(N_{\uparrow}) \eqdef N_0^\up $ and $u=\text{ceil}(N_{\uparrow})$. The minimum we obtain is

$\left(\Delta N_{\uparrow}\right)^{2}=\alpha_{\uparrow}\left(1-\alpha_{\uparrow}\right)$, where $\alpha_\s$ denotes the fractional part of $N_\s$.

\begin{figure}
\centering
\includegraphics[width=0.85\linewidth]{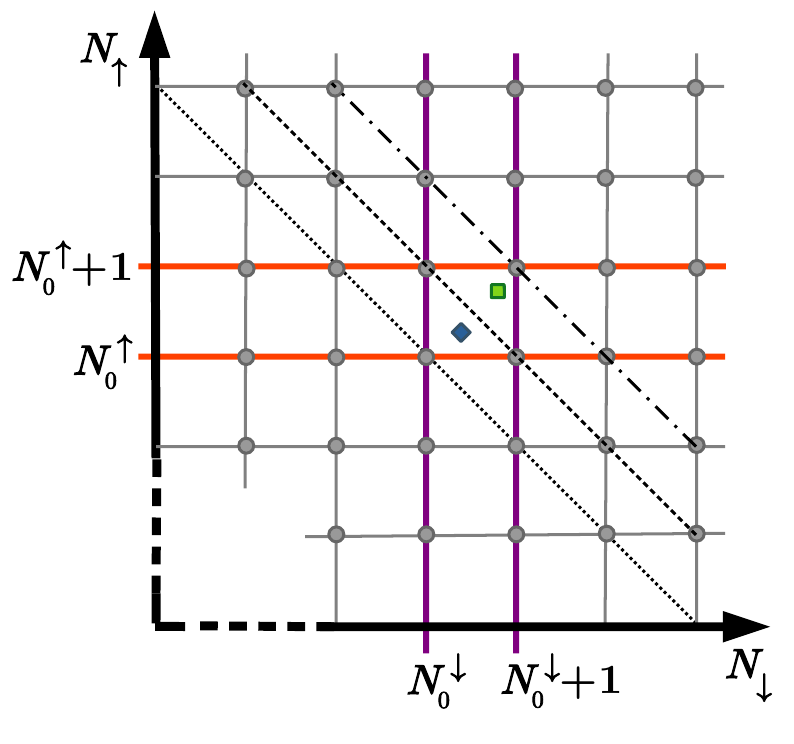}
\caption{Orange: the lines $N_\up = N_0^\up$ and $N_\up = N_0^\up+1$, which include all the points that may have nonzero weights, in order to minimize $(\Delta N_\up)^2$. Purple: the lines $N_\dw = N_0^\dw$ and $N_\dw = N_0^\dw+1$, which include all the points that may have nonzero weights, in order to minimize $(\Delta N_\dw)^2$. Green square: a point with fractional $N_\up$ and $N_\dw$, such that $\alpha_\up+\alpha_\dw>1$. Blue rhombus: a point with fractional $N_\up$ and $N_\dw$, such that $\alpha_\up+\alpha_\dw<1$. The lines $N = N_0^\up + N_0^\dw$ (dotted), $N = N_0^\up + N_0^\dw+1$ (dashed) and $N = N_0^\up + N_0^\dw+2$ (dash-dot) are added for comparison. They may correspond to $N=N_0-1$, $N=N_0$ and $N=N_0+1$ (case of green square) or to $N=N_0$, $N=N_0+1$ and $N=N_0+2$ (case of blue rhombus)}
    \label{fig:3point_DSz_illustration}
\end{figure}

The above logic applies identically to $\left(\Delta N_{\downarrow}\right)^{2}$: this quantity is minimized by assigning non-zero coefficients only to pure states with $n_\dw=\text{floor}(N_{\dw}) \eqdef N_0^\dw$ and $n_\dw= \text{ceil}(N_{\dw})$, resulting in $\left(\Delta N_{\downarrow}\right)^{2}=\alpha_\dw(1-\alpha_\dw)$.
Returning to the minimization of $\left(\Delta N_{\uparrow}\right)^{2}+\left(\Delta N_{\downarrow}\right)^{2}$, it is clear that the minimum possible value of this quantity is $\alpha_{\uparrow}\left(1-\alpha_{\uparrow}\right)+\alpha_{\downarrow}\left(1-\alpha_{\downarrow}\right)$. In Fig.~\ref{fig:3point_DSz_illustration} we illustrate the situation: in order to minimize $(\Delta N_\up)^2$, only weights residing along the orange lines should be included, and all the others are zero. Likewise, in order to minimize $(\Delta N_\dw)^2$, only weights residing along the purple lines should be included. Both of $(\Delta N_\s)^2$ are minimized simultaneously \emph{only} by an ensemble comprised of the four pure states, each residing at an intersection of a purple and an orange line: $(N_0^\up,N_0^\dw)$, $(N_0^\up+1,N_0^\dw)$, $(N_0^\up,N_0^\dw+1)$ and $(N_0^\up+1,N_0^\dw+1)$.

We notice that out of these four states, only three are allowed to participate in the ensemble -- those which have $N=N_{0}$ or $N=N_{0}+1$. The fourth state will have either $N=N_0-1$ or $N=N_0+2$. This fourth state, excluded from the ensemble, may be the one at $(N_0^\up,N_0^\dw)$, or $(N_0^\up+1,N_0^\dw+1)$, depending on the point $(N_\up,N_\dw)$, namely depending on whether $\alpha_\up+\alpha_\dw$ is greater than 1, or less than 1.

In Fig.~\ref{fig:3point_DSz_illustration} both these cases are shown: for the green square, representing a point $(N_\up,N_\dw)$ where $\alpha_\up+\alpha_\dw>1$, the point $(N_0^\up,N_0^\dw)$ is excluded from the ensemble. In contrast, the blue rhombus corresponds to some $(N_\up,N_\dw)$ where $\alpha_\up+\alpha_\dw<1$, and thus the point $(N_0^\up+1,N_0^\dw+1)$ is excluded.

Thus, the remaining three points form the unique ensemble, which minimizes $\Delta S_{z}$, resulting in $\left(\Delta S_{z}\right)^{2}=\frac{1}{2}\left(\alpha_{\uparrow}\left(1-\alpha_{\uparrow}\right)+\alpha_{\downarrow}\left(1-\alpha_{\downarrow}\right)\right)-\frac{1}{4}\alpha\left(1-\alpha\right)$.

\section{lemma on the maximum entropy ensemble}\label{sec:maxS}
In this Section, we consider an ensemble within Region~\EqSpinCond, $\hat{\Gamma}=\sum_\NM \gamma_\NM \ketbra{\Psi_\NM}$, satisfying the constraints \EqconstraintRange, ~\EqconstraintOne, ~\EqconstraintTwo ~and \EqconstraintThree ~for some $N_\tot$ and $M_\tot$, and aim to show that the $\hat{\Gamma}$ which maximizes the entropy $\mathscr{S}(\hat{\Gamma})$ includes all allowed pure states, i.e.\ those which may participate in $\hat{\Gamma}$ without violating the constraints. When $N_\tot$ is strictly fractional and $|M_\tot|<M_B$, the allowed pure states are \emph{all} those which appear in Eq.~\EqGammaTrapezoids.

We proceed to show a more general version of this statement, which applies also outside the context of the present work. Consider an ensemble
\begin{equation} \label{eq:gen_Gamma}
    \hat{\Gamma}=\sum_i \gamma_i \ketbra{\Psi_i}
    \end{equation}
    of orthonormal pure states $\ket{\Psi_i}$, bound by the constraints
\begin{equation}\label{eq:gen_lin_constraints}
\sum_i \gamma_i c_{ij}=d_j, \quad \gamma_i\in(0,1]
\end{equation}
for some constants $d_j$ and $c_{ij}$. In the context of the present work, the specific linear constraints are
Eqs.~\EqconstraintOne,~\EqconstraintTwo ~and \EqconstraintThree.
Assume that some set of orthonormal pure states $\{\ket{\Phi_k}\}$, which are all orthogonal to each $\ket{\Psi_i}$, and are not included in $\hat{\Gamma}$, can also take part in the ensemble. This means we have an ensemble
$\hat{\Xi}=\sum_k f_k\ketbra{\Phi_k}+\sum_i\xi_i \ketbra{\Psi_i}$ with $f_k\neq0$, which also satisfies the constraints of Eqs.~\eqref{eq:gen_lin_constraints} (where the coefficients $\xi_i$ and $f_k$ are appropriately substituted for $\gamma_i$).

We construct a new ensemble $\hat{\Gamma '}=(1-\varepsilon)\hat{\Gamma}+\varepsilon\hat{\Xi}$, which then satisfies Eqs.~\eqref{eq:gen_lin_constraints}, as well.
We now claim that for a sufficiently small $\varepsilon>0$, the entropy of the new ensemble satisfies $\mathscr{S}(\hat{\Gamma'})>\mathscr{S}(\hat{\Gamma})$.
Calculating this new entropy explicitly, we get
\begin{align}\label{eq:entropy_new_Sprime}
\nonumber
\mathscr{S}(\hat{\Gamma'}) &= -\sum_i ((1-\varepsilon)\gamma_i+\varepsilon \xi_i)\ln((1-\varepsilon)\gamma_i+\varepsilon \xi_i)- \\
&\quad -\sum_k \varepsilon f_k\ln(\varepsilon f_k).
\end{align}

In the limit of $\varepsilon\to0^+$, we notice that
\begin{equation}
    {((1-\varepsilon)\gamma_i+\varepsilon \xi_i)\ln((1-\varepsilon)\gamma_i+\varepsilon \xi_i)=\gamma_i\ln\gamma_i+O(\varepsilon)}
\end{equation}
(recall $\gamma_i\neq0$) and that $[-\varepsilon f_k\ln(\varepsilon f_k)]/\varepsilon \to \infty$. The latter means that the second sum in Eq.~\eqref{eq:entropy_new_Sprime} is positive and approaches $0$, but dominates over $O(\varepsilon)$.
Therefore, in this limit,
\begin{equation}
    {\mathscr{S}(\hat{\Gamma'})=\mathscr{S}(\hat{\Gamma})-\sum_{k}\varepsilon f_k \ln(\varepsilon f_k)+O(\varepsilon) > \mathscr{S}(\hat{\Gamma}}).
\end{equation}

This concludes the proof: by adding new pure states, with infinitesimal coefficients, to a given ensemble, we increased its entropy, without
violating the constraints. Therefore, when searching for maximal entropy, we have to consider only those ensembles, which include \emph{all} allowed states. Thus, the ensemble coefficients $\lambda_\NM$ which maximize the entropy are as derived in Sec.~II\thinspace A of the main text.

\section{Solution of Eq.~\SmaxMtotCondition} \label{sec:ContDiff_Eq39}
As we have seen in the main text, the coefficients of the maximum entropy ensemble are determined by the solution $z\in(0,\infty)$ to Eq.~\SmaxMtotCondition.
We shall now demonstrate the existence and uniqueness of this solution. Changing the variable to $t=\ln z$, the equation is of the form $f(t)=M_\tot$, where
\begin{align}
\nonumber
&f(t) = (1-\alpha)\Big(S_0+\tfrac12\Big)\,\coth\Big(\big(S_0+\tfrac12\big)t\Big)+ \\
     &\quad \alpha\Big(S_1+\tfrac12\Big)\,\coth\Big(\big(S_1+\tfrac12\big)t\Big)
       - \tfrac12\,\coth\Big(\tfrac{t}{2}\Big).
\end{align}

\begin{figure}[h]
    \centering
\includegraphics[width=0.95\linewidth]{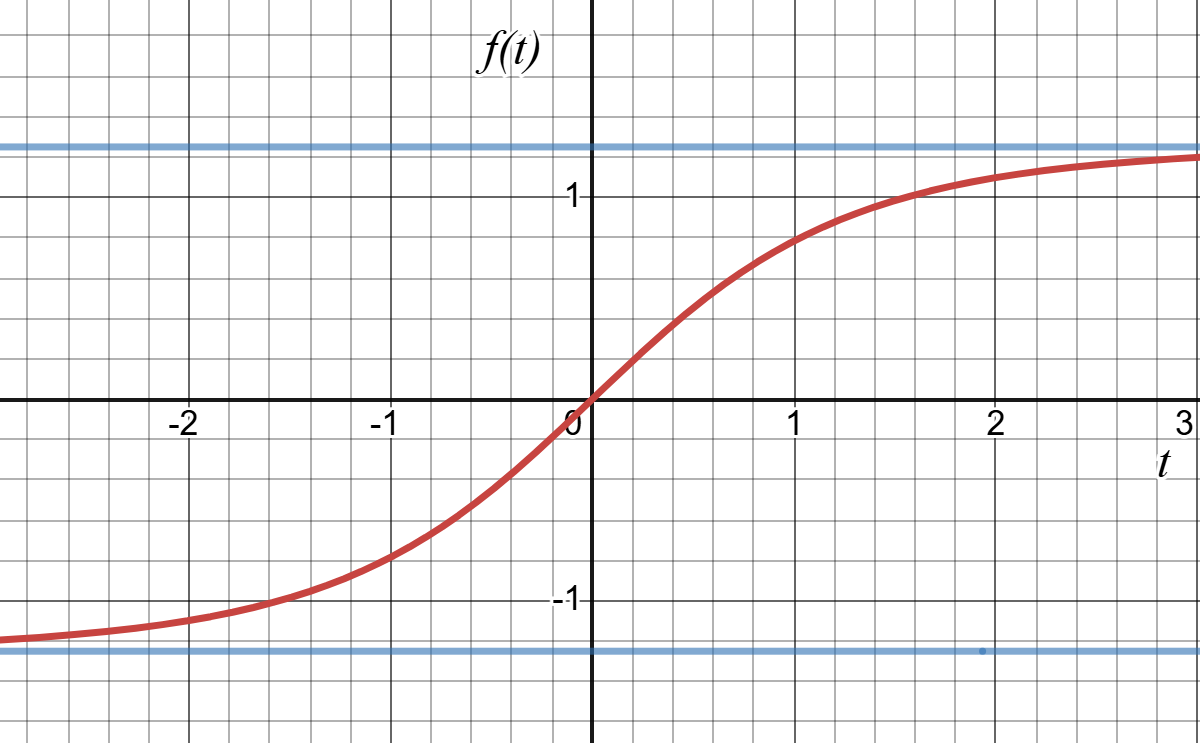}
\caption{Graph of $f(t)$ with the parameters $S_0=1$, $S_1=1.5$,  $\alpha=0.5$ and therefore $M_B=1.25$. The lines $f(t)=\pm M_B$ are shown in blue.}
    \label{fig:maxS_f}
\end{figure}

A graph of this function is shown in Fig.~\ref{fig:maxS_f}.  Since $\lim_{t\to-\infty}f(t)=-M_B$ and $\lim_{t\to \infty}f(t)=M_B$, the existence of a solution is guaranteed if $f(t)$ is continuous. If $f(t)$ is also strictly increasing, the solution is unique, as well.

It is clear that $f(t)$ is continuous at $t\neq0$. In the limit of $t\to0$, the function is also regular, approaching $0$: by expanding $\coth(x)=\frac{1}{x}+\frac{x}{3}+O(x^3)$ about $x=0$, we see that  \begin{equation}\label{eq:f_near_zero}
f\left(t\right)=\left(\frac{\left(1-\alpha\right)\left(S_{0}+\frac{1}{2}\right)^{2}+\alpha\left(S_{1}+\frac{1}{2}\right)^{2}}{3}-\frac{1}{12}\right)t+O\left(t^{3}\right).
\end{equation}
Thus, $f(t)$ is indeed continuous.
To show that $f(t)$ is strictly increasing, we differentiate
\begin{align}\label{eq:dfdt}
    \nonumber&\frac{df}{dt}=\frac{\left(\frac{1}{2}\right)^{2}}{\sinh^{2}\left(\frac{t}{2}\right)}-\left(1-\alpha\right)\frac{\left(S_{0}+\frac{1}{2}\right)^{2}}{\sinh^{2}\left(\left(S_{0}+\frac{1}{2}\right)t\right)} -
    \\
&\alpha\frac{\left(S_{1}+\frac{1}{2}\right)^{2}}{\sinh^{2}\left(\left(S_{1}+\frac{1}{2}\right)t\right)}.
\end{align}

Next, we notice that the expression $\left(\frac{w}{\sinh\left(wt\right)}\right)^2$ is a strictly decreasing function of $w$, at $w>0$ and $t\neq0$. As a result, $\left(\frac{\thalf}{\sinh\left(\frac{t}{2}\right)}\right)^2\geqs\left(\frac{S_{0,1}+\thalf}{\sinh\left(\left(S_{0,1}+\thalf\right)t\right)}\right)^2$, and this inequality is strict if $S_{0,1}\neq 0$ (which always holds for either $S_0$ or $S_1$). Substituting these into Eq.~\eqref{eq:dfdt}, we have $\frac{df}{dt}\geqs0$ for all $t\neq0$.

Likewise for $t=0$: from Eq.~\eqref{eq:f_near_zero} the derivative near $t=0$ is $\frac{df}{dt}=\frac{1}{3} \lp[\left(1-\alpha\right)\left(S_{0}+\frac{1}{2}\right)^{2}+\alpha\left(S_{1}+\frac{1}{2}\right)^{2} \rp] -\frac{1}{12}+O\left(t^{2}\right)$.
 In particular, the derivative is well-behaved and non-negative at $t=0$.
Thus, for any $t\in\mathbb{R}$, the equality $\frac{df}{dt}=0$ occurs only if $\alpha=0$ and $S_0=0$, in which case there is no ambiguity of the ground state in the first place. Otherwise,  $\frac{df}{dt}>0$, hence $f(t)$ is strictly increasing.

To conclude, whenever there is an ambiguity to be removed, $f(t)$ is both continuous and strictly increasing as required, meaning Eq.~\SmaxMtotCondition ~has precisely one solution.

\section{Maximum entropy ensemble in presence of internal degeneracy} \label{sec:InternalDeg}

In this Section, we introduce an internal degeneracy of the pure states at each point $(N,M)$, and generalize the maximal-entropy ground state within Region~\EqSpinCond~to this scenario. This internal degeneracy is assumed to not be accidental, but arise from symmetries of the system, generated by Hermitian operators $\hat{A}_1,\ldots,\hat{A}_t$ which commute with each other, with $\hat{H}$, $\hat{N}$ and with $\hat{S}_z$, to form a complete set of commuting observables. We define the \emph{tuple} of these operators, $\hat{A}=(\hat{A}_1,\ldots,\hat{A}_t)$.
To derive the maximum entropy in similar fashion to Sec.~\SecRegionOne~of the MT, we must further assume that $\hat{A}$  commutes with the spin ladder operator $\hat{S}_+ =\hat{S}_x+i\hat{S}_y$.
Typically, it is indeed the case that $[\hat{A},\hat{\mathbf{S}}]=0$ and therefore $[\hat{A},\hat{S}_+]=0$, for example if the symmetry of the system is rotation or reflection.

The pure ground states at each $(N,M)$ are then labeled $\ket{\Psi_{\NM}^i}$, having eigenvalues $N$, $M$ and $a_i = (a^1_i, \ldots, a^t_i)$, with respect to the operators $\hat{N}$, $\hat{S}_z$ and $\hat{A}$. Here $i=1,\ldots, g_\NM$ indexes the degenerate pure states, and $g_\NM$ is the degree of the degeneracy at the point $(N,M)$.

Since $\hat{A}$ commutes with $\hat{S}_{+}$, whose action raises $M$ by~1, the state $\hat{S}_+ \ket{\Psi_{\NM}^i}$ is an eigenstate of $\hat{A}$ with eigenvalue $a_i$, for each $(N,M)$ and each $i$.
Consequently, $\hat{S}_+$ preserves the degeneracy of the pure ground states at $(N,M)$, meaning that $g_{\NM}$ is independent of $M$. We therefore denote $g_N=g_{\NM}$. This argument is in similar fashion to our discussion of $M$-independent quantities (see Eqs.~\EqPPWL--\EqPNMminusone~and their pertinent discussion).

The ensemble at some $(N_\tot,M_\tot)$ within Region~\EqSpinCond~ is then defined by statistical weights $\lambda_\NM^i$ assigned to each $\ket{\Psi_\NM^i}$.
The entropy is therefore expressed as
\begin{equation}\label{eq:entropy_degen_def}
    \mathscr{S}=-\sum_{\NM}\sum_{i=1}^{g_{N}} \lambda_{\NM}^i \ln \lambda_\NM^i .
\end{equation}
We define the total weight of each $(N,M)$ point as
\begin{equation}\label{eq:lambda_tilde_degen}
\tilde{\lambda}_\NM=\sum_{i=1}^{g_N} \lambda_{\NM}^i.
\end{equation}

 We proceed by first maximizing the entropy of Eq.~\eqref{eq:entropy_degen_def} with respect to $\{\lambda_\NM^i\}$, with Eq.~\eqref{eq:lambda_tilde_degen} as our constraint (together with $\lambda_\NM^i \in[0,1]$).
In this particular case, we can maximize each term  $-\sum_{i=1}^{g_N} \lambda_\NM ^i \ln \lambda_\NM ^i$ separately,  while satisfying  Eq.~\eqref{eq:lambda_tilde_degen}, for each $(N,M)$. Performing a maximization procedure with a Lagrange multiplier, similarly to what we described in other contexts above, we find that for each $(N,M)$ all weights must equal each other, i.e.
\begin{equation}\label{eq:degen_lambda_i_equal}
     \lambda_\NM^i=\frac{\tilde{\lambda}_\NM}{g_N}.
 \end{equation}
This is a standard outcome of entropy maximization.
Therefore, Eq.~\eqref{eq:entropy_degen_def} becomes
\begin{align}
  \mathscr{S} &=-\sum_{\NM}\sum_{i=1}^{g_{N}}\frac{\tilde{\lambda}_{\NM}}{g_{N}}\ln\frac{\tilde{\lambda}_{\NM}}{g_{N}}= \nonumber \\
  &= -\sum_{\NM}\tilde{\lambda}_{\NM}\ln\tilde{\lambda}_{\NM}+\sum_{\NM}\tilde{\lambda}_{\NM}\ln g_{N}.
\end{align}
Due to the constraints imposed by Eqs.~\EqconstraintOne~ and \EqconstraintTwo, the $N=N_0$ and $N_0+1$ weights must sum to $1-\alpha$ and $\alpha$, respectively:
$\sum_{M=-S_0}^{S_0}\tilde{\lambda}_{N_0 M}=1-\alpha$ and $\sum_{M=-S_1}^{S_1}\tilde{\lambda}_{N_0+1 M}=\alpha$.  Substituting these into our expression for the entropy and relying on the fact that the degeneracy degree $g_N$ is $M$-independent, we find that
\begin{equation}\label{eq:entropy_degen_simplified}
    \mathscr{S}=-\sum_{\NM}\tilde{\lambda}_{\NM}\ln\tilde{\lambda}_{\NM}+\left(1-\alpha\right)\ln g_{N_{0}}+\alpha\ln g_{N_{0}+1}.
\end{equation}

Our remaining task is to maximize the entropy of Eq.~\eqref{eq:entropy_degen_simplified}. We realize that the second and third terms in the equation are constants, and are not subject to the maximization.
Thus, this is precisely the same maximization we have already encountered in Eq.~\EqDefEntropy, where the pure ground states were not degenerate, i.e. $g_{N_0}=g_{N_0+1}=1$.
Consequently, $\tilde{\lambda}_\NM$ are given by Eqs.~\SmaxMtotCondition~ and
\EqMaxSLambdas.

We conclude that the total weights $\tilde{\lambda}_\NM$ are the same as those we have already derived in the absence of internal degeneracy. For maximal entropy, at each $(N,M)$, these weights should be equally divided among all of the degenerate pure states $\ket{\Psi_\NM^i}$, as in Eq.~\eqref{eq:degen_lambda_i_equal}.
\bibliography{bib2023}